\def\bold#1{\setbox0=\hbox{$#1$}%
     \kern-.025em\copy0\kern-\wd0
     \kern.05em\copy0\kern-\wd0
     \kern-.025em\raise.0433em\box0 }

\def\slash#1{\setbox0=\hbox{$#1$}#1\hskip-\wd0\dimen0=5pt\advance
       \dimen0 by-\ht0\advance\dimen0 by\dp0\lower0.5\dimen0\hbox
         to\wd0{\hss\sl/\/\hss}}

\documentstyle[12pt]{article}

\newlength{\dinwidth}
\newlength{\dinmargin}
\newcommand{\resection}[1]{\setcounter{equation}{0}\section{#1}}

\def\today{\ifcase\month
\or January\or February\or March\or April\or May
\or June\or July\or August\or September\or October
\or November \or December\fi \space\number\day, \number\year}
\def\draft{\hfill{\it File:\jobname, Draft:\today}}

\begin{document}
\vspace*{5cm}
\begin{center}
  \begin{Large}
  \begin{bf}
THE CHIRAL APPROACH TO THE ELECTROWEAK INTERACTIONS$^*$\\
  \end{bf}
  \end{Large}
  \vspace{1.5cm}
  \begin{large}
F. Feruglio\\
  \end{large}
Dipartimento di Fisica, Univ.
di Padova\\
I.N.F.N., Sezione di Padova\\
  \vspace{1.5cm}
\end{center}
\begin{quotation}
\begin{center}
  \begin{Large}
  \begin{bf}
  ABSTRACT
  \end{bf}
  \end{Large}
\end{center}
  \vspace{5mm}
\noindent
The effective lagrangian approach is reviewed and applied to the case
of electroweak interactions.
\end{quotation}

  \vspace{1.5cm}
\begin{center}
DFPD 92/TH/50\\
September 1992\\
\vspace{1cm}
\noindent
$^*$ {\it Lectures given at the\\
$2^{nd}$ National Seminar of Theoretical Physics\\
Parma, 1-12 September 1992.}
\end{center}
\newpage
\thispagestyle{empty}
\vspace*{7cm}
\begin{quotation}
\tableofcontents
\end{quotation}

\newpage
\setcounter{page}{1}
\def\lq{\left [}
\def\rq{\right ]}
\def\LL{{\cal L}}
\def\DD{{\cal D}}
\def\dmu{{\partial_\mu}}
\def\dnu{{\partial_\nu}}
\def\dmua{{\partial^\mu}}
\def\ss{{\left(
         \begin{array}{cc}
         \sigma&0\nn\\
         0&\sigma
         \end{array}
         \right)}}
\def\sv{{
         \left(
         \begin{array}{c}
         0\nn\\
         \sigma
         \end{array}
         \right)
         }}
\def\svu{{
         \left(
         \begin{array}{c}
         \sigma\nn\\
         0
         \end{array}
         \right)
         }}
\def\gl{{e^{\dd i\vec\alpha\cdot\frac{\vec\tau}{2}}}}
\def\glm{{e^{\dd -i\vec\alpha\cdot\frac{\vec\tau}{2}}}}
\def\gr{{e^{\dd i\alpha_Y\frac{\tau^3}{2}}}}
\def\grm{{e^{\dd -i\alpha_Y\frac{\tau^3}{2}}}}
\def\da{\frac{\delta\alpha}{\alpha}}
\def\dg{\frac{\delta G_F}{G_F}}
\def\dmz{\frac{\delta M_Z^2}{M_Z^2}}
\def\dmw{\frac{\delta M_W^2}{M_W^2}}

\newcommand{\be}{\begin{equation}}
\newcommand{\ee}{\end{equation}}
\newcommand{\bea}{\begin{eqnarray}}
\newcommand{\eea}{\end{eqnarray}}
\newcommand{\nn}{\nonumber}
\newcommand{\dd}{\displaystyle}

\resection{Introduction}
\indent
The aim of this course is to provide
an introduction to the basic features and methods
of effective field theories, working out, as an example,
the case of electroweak interactions.

The use of effective lagrangians dates back to the early 60's
with the introduction of the non-linear $\sigma$-model \cite{Gursey} as
an effective model for low energy strong interactions, explicitly
exhibiting chiral symmetry breakdown.
The theoretical basis of chiral lagrangians was later formulated by Weinberg
\cite{Weinberg} in an attempt to characterize the most general
S-matrix elements
compatible with chiral symmetry and the general requirements
of analyticity, unitarity and cluster decomposition property.

Chiral lagrangians are essentially tailored to describe the phenomenon
of spontaneous symmetry breaking, which plays a major role in both
strong and electroweak interactions. They can be regarded as
the low energy limit of an underlying fundamental theory.
In the case of strong interactions such a regime is difficult to discuss
in the framework of QCD since the knowledge of the non-perturbative
dynamical effects is required. Therefore, it is not surprising that
in the context of strong interactions the chiral lagrangian approach
has found its greatest development \cite{Gasser,Leut,Ecker}.

More recently, effective field theories have received attention
in the analysis of radiative corrections for electroweak theories
\cite{Terning,Golden,Georgi2,Holdom,DeRujula}.
Indeed, in the standard model of electroweak interactions,
low energy effective lagrangians naturally
occur when some of the particles of the theory become very heavy.
Much interest has been devoted to the possibility of an heavy Higgs
\cite{Bernard,Appelquist,Longhi},
which is central in the discussion of the electroweak symmetry
breaking itself \cite{Veltman}.

Interesting problems, related to the non-applicability
of the decoupling theorem \cite{Carazzone},
are also raised by the possibility of heavy fermions
\cite{D'Hoker,Steger,me,D'Hoker2}.
In particular, the case of an heavy top quark calls for the use of the whole
apparatus of non-linear realizations, the usual linear realization of the
electroweak symmetry being destroyed from the beginning \cite{me}. By removing
heavy fermions from the low energy spectrum, one has also to take care
of possible anomalies in the gauge currents \cite{Anomaly}.
These anomalies are consistently avoided
thanks to the presence of a Wess-Zumino term \cite{WZ} in the effective action
and to the very specific interplay of the latter
with the non-linear realization of the symmetry.

Another fruitful application is that when the fundamental theory
one is dealing with is not known, or not precisely defined (actually,
this was the case with the non-linear $\sigma$-model).
If the possible symmetries are known, then chiral lagrangians may be
naturally introduced. This happens, for instance, with technicolor theories
\cite{Technicolor},
where a considerable amount of information can be extracted from
the analysis of the corresponding low energy models
\cite{Chadha,Terning,Golden}.
Chiral lagrangians are characterized by an infinite tower
of non-renormalizable operators, actually all the operators which
are consistent with the assumed symmetries. To make practical use of
such an infinite expansion one has to be able to select the
most relevant terms. In some cases,
when the full underlying theory is given, the size of the possible terms
can be computed, at least in principle, by matching the predictions
of the fundamental field theory and those of the effective one,
at some reference
scale \cite{Georgi3}. This procedure, however, may not be always viable.
In this case, some insight about
the relevance of the various terms can be obtained by a judicious use
of dimensional analysis \cite{Manohar}.

An interesting development of the effective lagrangian approach
was offered by the study of the hidden gauge symmetries \cite {Hidden}
possessed by chiral models. Applications of these symmetries
have been made in low energy strong interactions, to describe the
lowest spin 1 states \cite{Bando},
and in the study of the so-called strong electroweak sector,
to analyze the effects of possible vector resonances \cite {Casalbuoni}.
The strongly interacting symmetry breaking sector has been
extensively studied by means of chiral lagrangian techniques
\cite{Herrero}.

Effective field theories can also be introduced to perform
a model independent analysis of $SU(2)_L\otimes U(1)_Y$
supersymmetric gauge theories \cite{Porrati}.
Finally, the case of N=1 supergravity \cite{sugra}
shows a certain analogy with the
examples quoted above, if interpreted as an effective low energy theory
derived by integrating out the massive modes of superstring theories
\cite{superstring}.

The lectures are organized as follows. Section 2 contains a review
of the non-linear realizations of a Lie group. The formalism is
then applied to the case of weak interactions, whose lowest order
lagrangian is derived in section 3. Section 4 presents a
simplified discussion of the electroweak radiative corrections,
focussing on the LEP I physics. In section 5, the effective electroweak
lagrangian is extended to account for quantum effects from heavy particles,
and the physical content of the theory is detailed in section 6.
Finally section 7 provides an example of matching between low energy
and high energy physics, leading to a specification of the
parameters in the effective lagrangian.

\newpage

\resection{Non-Linear Realizations}

This section contains a short review of the non-linear realizations
of a Lie group $G$ \cite{Callan}. We consider a real analytic manifold $M$,
together with a Lie Group $G$
\footnote{compact, connected and semisimple}
of transformations acting on $M$:
\be
x\to gx~~~~~~~~x\in M,~~~~~g\in G
\ee
We assume that there is a special point of $M$ called origin, described by
the null vector $0$ and invariant under the action of a continuous
subgroup $H$ of $G$:
\be
h0=0~~~~~~~~h\in H
\ee

The physical situation one has in mind is that of a manifold of
scalar fields
with the origin describing the vacuum configuration. The group $G$
is the (global or local) invariance group of the theory and the
subgroup $H$ is the invariance group of the vacuum. In other words,
one is dealing with the spontaneous breaking of $G$ into $H$.

Our purpose is to characterize all possible
non-linear realizations of the group $G$ on the manifold $M$,
that is to classify all possible theories corresponding
to that pattern of symmetry breaking.

Preliminary to the analysis of this problem is a general result of
quantum field theory establishing the independence of the physical
content of a theory from the choice of interpolating fields \cite{Haag}.
In
the framework of a lagrangian formulation of the theory, the theorem
can be rephrased in the following form \cite{Coleman}.

\noindent
$\bullet$
{\bf Theorem}: If a theory is defined by the lagrangian density:
\be
\LL=\LL(\phi,\dmu\phi)
\ee
depending on a set of scalar fields $\phi$,
and if the following local transformation of fields is performed:
\be
\phi=F(\phi')~~,
\ee
then the transformed lagrangian density:
\be
\LL'(\phi',\dmu\phi')=\LL(F(\phi'),\dmu F(\phi'))
\ee
defines in general a new theory. Nevertheless, provided the transformation
(2.4)
has a Jacobian determinant equal to one at the origin, the S-matrix elements
of the two theories are the same.

{}From now on, we will refer to such
transformations as allowed ones. The above theorem also holds order by order
in perturbation theory.
\vskip 1.truecm

Another useful result is that
the more general non-linear realization of the group $G$
can be regarded as linear, once specialized to the subgroup $H$ of $G$.

\noindent
$\bullet$
{\bf Theorem}: If $H$ is the subgroup of $G$ leaving the origin invariant,
         then it is always possible to choose coordinates on $M$
         so that:
\be
hy=D(h)y~~~~~~~\forall h\in H
\ee
where $D(h)$ is a linear representation of $H$.

We sketch below the proof of this theorem.
By expanding the element $hx$ in powers of $x$ around the origin one has:
\be
hx=D(h)x+O(x^2)
\ee
In eq. (2.7), $O(x^2)$ denotes terms which are at least quadratic in $x$.
The constant term is absent, due to the fact that $H$ leaves
the origin invariant.
We consider an invariant measure $dh$ on the group $H$, normalized so that:
\be
\int_H dh=1~~,
\ee
and we define the following coordinates on $M$:
\bea
y&=&\int_H dh D^{-1}(h) hx\nn\\
 &=&x+O(x^2)
\eea
Acting with an element $h_0$ of $H$ on the point $y$, one obtains:
\bea
h_0y&=&\int_H dh D^{-1}(h) hh_0x\nn\\
    &=&\int_H d(hh_0) D^{-1}(hh_0h_0^{-1}) (hh_0)x\nn\\
    &=&D(h_0)y
\eea
Notice that the transformation given in eq. (2.9) is an allowed one.

We now give an example of a non-linear realization of $G$ becoming
linear when it is restricted to the subgroup $H$. As we shall see,
this example plays a central role in our discussion and the non-linear
realization
dealt with is said to be in the standard form.
To start with, one introduces a complete set of generators of $G$, $(V_i,A_l)$,
orthonormal with respect to the inner Cartan product and such that the $V$'s
are generators of the subgroup $H$. Each element $g_0$ of $G$ admits the
unique decomposition:
\be
g_0=e^{\dd\xi A}e^{\dd u V}
\ee
where $\xi A=\xi_l A_l$ and $u V=u_i V_i$.
For every element $g\in G$, one has:
\be
ge^{\dd\xi A}=e^{\dd\xi^g A}e^{\dd u^g V}
\ee
with
\bea
\Biggl\{
\begin{array}{ccc}
\xi^g&=&\xi^g(\xi)\\
u^g&=&u^g(\xi)
\end{array}
\eea
The idea is to use $\xi$, the parameters related to the $A$ generators,
as a subset of coordinates of $M$, with a transformation law under $G$ defined
by eq. (2.13). To complete the set of coordinates, one introduces a vector
$\psi$ carrying a linear representation of $H$:
\be
\psi\to D(h)\psi~~~~~~~~h\in H
\ee
It is not difficult to show that the transformations
\bea
\Biggl\{
\begin{array}{ccc}
\xi&\to&\xi^g(\xi)\\
\psi&\to&D(e^{\dd u^g(\xi) V})\psi
\end{array}
\eea
provide a non-linear realization of the group $G$, the so-called standard one.
By restricting the transformations to $H$, one has:
\be
he^{\dd\xi A}=he^{\dd\xi A}h^{-1}h
\ee
from which one deduces:
\bea
\Biggl\{
\begin{array}{ccc}
e^{\dd\xi A}&\to&he^{\dd\xi A}h^{-1}\\
\psi&\to&D(h)\psi
\end{array}
\eea
Therefore,
the transformations of $H$ on the standard coordinates are linearly realized.

The main result, not proved here, is contained in the following statement.

\noindent
$\bullet$
{\bf Theorem}: any non-linear realization of G can be put into the standard
form
by an allowed coordinate transformation.

This theorem solves the problem of characterizing all possible non-linear
realization of $G$ on $M$.
Its physical content is that, in discussing a theory
describing the spontaneous breaking of the Lie group $G$ into a subgroup $H$,
it is not restrictive to choose a set of fields transforming according to
the standard form given in eq. (2.15).

\noindent
$\bullet$
{\bf Example}: we choose $G=SU(2)_L\otimes SU(2)_R$ and $H$ its diagonal
subgroup, $SU(2)_{L+R}$.
The generators of $G$, $L_i$ and $R_i~(i=1,2,3)$, satisfy
the following algebra:
\bea
{[L_i,L_j]}&=&i\epsilon_{ijk} L_k\nn\\
{[R_i,R_j]}&=&i\epsilon_{ijk} R_k\nn\\
{[L_i,R_j]}&=&0
\eea
As a realization of this algebra, one can take the following $4\times 4$
matrices:
\be
{L_i}=
\left(
\begin{array}{cc}
\dd\frac{\tau^i}{2}&0\nn\\
0&0
\end{array}
\right)
{}~~~~~
{R_i}=
\left(
\begin{array}{cc}
0&0\nn\\
0&\dd\frac{\tau^i}{2}
\end{array}
\right)
\ee
where $\tau^i$ are the Pauli matrices.
The generators $(V,A)$ are given by:
\bea
V_i&=&L_i+R_i\nn\\
A_i&=&L_i-R_i
\eea
and satisfy the algebra:
\bea
{[V_i,V_j]}&=&i\epsilon_{ijk} V_k\nn\\
{[V_i,A_j]}&=&i\epsilon_{ijk} A_k\nn\\
{[A_i,A_j]}&=&i\epsilon_{ijk} V_k
\eea
To every generator $A_i$ we associate a coordinate $\xi_i$, with the
transformation law given by:
\bea
g e^{\dd i\xi A}&=&e^{\dd i\alpha A} e^{\dd i\beta V} e^{\dd i\xi A}\nn\\
            &=&e^{\dd i\xi^g A} e^{\dd i u^g V}
\eea
On the other hand, the group $G$ possesses an automorphism $g\to R(g)$,
such that:
\bea
V_i&\to&V_i\nn\\
A_i&\to&-A_i
\eea
and we can write:
\be
R(g) e^{\dd -i\xi A}=e^{\dd -i\xi^g A} e^{\dd i u^g V}
\ee
By combining eqs. (2.22) and (2.24), one finally obtains:
\be
e^{\dd 2i\xi^g A}=ge^{\dd 2i\xi A}R(g^{-1})
\ee
or, more explicitly:
\be
e^{\dd 2i\xi^g A}=e^{\dd i\alpha A}e^{\dd i\beta V}e^{\dd 2i\xi A}
e^{\dd -i\beta V}
e^{\dd i\alpha A}
\ee
{}From the last equality we can immediately see that,
by specifying the transformations
to the subgroup $H$, that is by putting the parameters $\alpha$ to zero,
one has linear transformations for the coordinates $\xi$.
\vskip 1.truecm

At this point we have to face the problem of constructing an invariant
formalism
out of the building block defined in eq. (2.15), the set of standard
coordinates and their transformation properties. After promoting the
coordinates on the manifold $M$ to scalar fields depending on the
space-time points, we realize that, even in the case of global
symmetry, the transformation laws given in eq. (2.15)
are local, because of the explicit dependence on the fields $\xi$'s. In
order
to work with objects depending on the derivatives of $\xi$ and $\psi$,
with simple transformation properties under the
group $G$, one can define appropriate covariant derivatives \cite{Callan},
dealing directly with the non-linearly transforming fields $\xi$ and $\psi$.
However, there is another possibility consisting in building functions of the
fields $(\xi,\psi)$ which transform linearly under the
group $G$. By combining such functions with the usual rules of
representation theory, it is then straightforward to define
invariant lagrangians. We will now show how to implement such a procedure.

Consider a linear representation $\DD(g)$ of $G$ containing in its
decomposition the representation $D(h)$ (see eqs. (2.14) and (2.15)).
We define:
\be
\Phi=\DD(e^{\dd\xi A})\psi~~.
\ee
The combination $\Phi$ given above transforms linearly under $G$,
according to the representation $\DD(g)$:
\bea
\Phi'&=&\DD(e^{\dd\xi^g A})D(e^{\dd u^g V})\psi\nn\\
     &=&\DD(g e^{\dd\xi A}e^{\dd -u^g V})D(e^{\dd u^g V})\psi\nn\\
     &=&\DD(g)\DD(e^{\dd\xi A})\psi\nn\\
     &=&\DD(g)\Phi
\eea

\noindent
$\bullet$
{\bf Example}: we take $G=SU(2)_L\otimes SU(2)_R$ and
$H=SU(2)_{L+R}$. Let $\sigma$
be an $SU(2)_{L+R}$-singlet, embedded into an $SU(2)_L\otimes SU(2)_R$
bidoublet:
\be
{\psi}=
\left(
\begin{array}{cc}
\sigma&0\nn\\
0&\sigma
\end{array}
\right)~~.
\ee
The action of the representation $\DD(g)$ on a bidoublet $\Omega$ is
defined as follows:
\bea
\Omega'&=&\DD(g)\Omega\nn\\
       &=&e^{\dd i \vec\lambda\cdot\frac{\vec \tau}{2}}
          \Omega
          e^{\dd -i \vec\rho\cdot\frac{\vec \tau}{2}}
\eea
where $\vec\lambda$ and $\vec\rho$ are the parameters of the
$SU(2)_L\otimes SU(2)_R$ transformations.
It is easy to verify that $\psi$ is a singlet under the subgroup
$SU(2)_{L+R}$, characterized by $\vec\lambda=\vec\rho$.
According to eq. (2.27), we now define:
\bea
\Phi&=&\DD(e^{\dd i\xi A})\psi\nn\\
    &=&e^{\dd i \vec\xi\cdot\frac{\vec \tau}{2}}\ss
          e^{\dd i \vec\xi\cdot\frac{\vec \tau}{2}}\nn\\
    &=&\sigma  e^{\dd i \vec\xi\cdot\vec\tau}
\eea
By construction, the function $\Phi=\Phi(\xi,\sigma)$ transforms as the
bidoublet $\Omega$ of eq. (2.30). Explicitly we have:
\bea
\sigma'&=&\sigma\\
e^{\dd i \vec\xi'\cdot\vec \tau}&=&e^{\dd i \vec\lambda\cdot\frac{\vec
\tau}{2}}
          e^{\dd i \vec\xi\cdot\vec \tau}
          e^{\dd -i \vec\rho\cdot\frac{\vec \tau}{2}}
\eea
To take into account the dimension of the fields $\xi$, one usually performs
the
following rescaling:
\be
\xi_i\to\frac{\xi_i}{f}
\ee
where $f$ is a constant with the dimension of a mass.
A lagrangian invariant under global transformations of $SU(2)_L\otimes SU(2)_R$
is:
\bea
\LL&=&\frac{f^2}{4}tr(\dmu U^\dagger \dmua U)\nn\\
   &=&\frac{1}{2}\dmu\xi_i\dmua\xi_i+...
\eea
where:
\be
U=e^{\dd i \frac{\vec\xi\cdot\vec \tau}{f}}
\ee
This is the lagrangian of the well-known non-linear $\sigma$-model
\cite {Gursey}. The global invariance under $SU(2)_L\otimes SU(2)_R$
is spontaneously broken down to $SU(2)_{L+R}$. The fields $\xi_i$,
associated to the broken generators $A_i$, are the Goldstone bosons.
The dots in eq. (2.35) denote higher order terms in the Goldstone fields.

\noindent
$\bullet$
{\bf Exercise}: verify that the transformation law for $U$ (see eq.(2.33)),
agrees with the
transformation law for $\xi$ given in eq. (2.26).

\noindent
$\bullet$
{\bf Exercise}: build the gauged non-linear $\sigma$-model by gauging the
entire
group $G=SU(2)_L\otimes SU(2)_R$. By going to the unitary gauge,
discuss the physical degrees of freedom and their mass spectrum.

\newpage
\resection{Electroweak Interactions: the Lowest Order}

In this section we review the construction of the standard model
of electroweak interactions and of some of its variants, by explicitly
applying the formalism developed in the previous lecture.
These models are characterized by an invariance under the gauge
group $G=SU(2)_L\otimes U(1)_Y$, spontaneously broken down to the
local subgroup $H=U(1)_{em}$. The generators $T^i$ ($i=1,2,3$) and $Y$
of $SU(2)_L\otimes U(1)_Y$ satisfy ~the following algebra:
\bea
{[T_i,T_j]}&=&i\epsilon_{ijk} T_k\nn\\
{[T_i,Y]}&=&0\nn\\
{[Y,Y]}&=&0
\eea
The generators of the subgroup $U(1)_{em}$ and of the coset
$SU(2)_L\otimes U(1)_Y/U(1)_{em}$ are given by
\footnote{different choices for the generators of the
coset are also possible}:
\bea
X_{em}&=&(T^3+Y)\longleftrightarrow U(1)_{em}\nn\\
Y^i   &=&T^i  ~~~~~~~~~\longleftrightarrow SU(2)_L\otimes U(1)_Y/U(1)_{em}
\eea
We begin by introducing the would-be Goldstone bosons and the Higgs field.
We consider a singlet $\sigma$ under $U(1)_{em}$, embedded into a
doublet representation $\DD$, of hypercharge $Y=1/2$:
\be
{\varphi}=
\left(
\begin{array}{c}
0\nn\\
\sigma
\end{array}
\right)
\ee
The action of the representation $\DD(g)$ on a doublet $\chi$ is defined
below:
\bea
\chi'&=&\DD(g)\chi\nn\\
     &=&e^{\dd i\vec\alpha\cdot\frac{\vec\tau}{2}}
        e^{\dd\alpha_Y\frac{1}{2}}
        \chi
\eea
In the equation above, $\alpha_i$ and $\alpha_Y$ are the local
parameters of the $SU(2)_L$ and $U(1)_Y$ gauge transformations,
respectively.
It is immediate to verify that $\varphi$ is a singlet under the
unbroken group $U(1)_{em}$. Following the prescription given
in eq. (2.27), we define:
\bea
\Phi&=&\DD(e^{\dd i\vec\xi\cdot\vec Y})\varphi\nn\\
    &=&e^{\dd i\vec\xi\cdot\frac{\vec\tau}{2}}{\sv}\\
\eea
By construction $\Phi$ transforms as a doublet with hypercharge
$Y=1/2$. We prefer to work with a linear multiplet written
in a matrix form and, to this purpose, we introduce the new doublet:
\bea
\tilde\Phi&=&i\tau^2\Phi^*\nn\\
             &=&e^{\dd i\vec\xi\cdot\frac{\vec\tau}{2}}{\svu}\\
\eea
The doublet $\tilde\Phi$ has hypercharge $Y=-1/2$. We define the
$2\times 2$ matrix $M$:
\bea
M&=&\left(\tilde\Phi\Phi\right)\nn\\
&=&\sigma e^{\dd i\vec\xi\cdot\frac{\vec\tau}{2}}
\eea
The matrix $M$ transforms as follows:
\be
M'=g_L M g_R^\dagger
\ee
where:
\bea
g_L&=&e^{\dd i\vec\alpha\cdot\frac{\vec\tau}{2}}\\
g_R&=&e^{\dd i\alpha_Y\frac{\tau^3}{2}}
\eea
Explicitly, one has:
\bea
\sigma'&=&\sigma\\
e^{\dd i\vec\xi'\cdot\frac{\vec\tau}{2}}&=&
g_L
e^{\dd i\vec\xi\cdot\frac{\vec\tau}{2}}
g_R^\dagger
\eea
The Higgs field, $\sigma$, is invariant under the whole gauge group
$SU(2)_L\otimes U(1)_Y$. The would-be Goldstone bosons transform
non-linearly under $SU(2)_L\otimes U(1)_Y$. However, the combination:
\be
U=e^{\dd i\frac{\vec\xi\cdot\vec\tau}{v}}
\ee
transforms linearly as specified in eq. (3.12). In the previous equality
we have rescaled the fields $\xi$, introducing the constant $v$.
Let us forget
for a while the Higgs field $\sigma$ and proceed to define
the lagrangian for the bosonic fields. The covariant derivative
for the combination $U$ is defined as follows:
\be
D_\mu U=\dmu U - g \hat W_\mu U+g'U\hat B_\mu
\ee
where $g$ and $g'$ are the gauge coupling constants for $SU(2)_L$ and
$U(1)_Y$; $\hat W_\mu,~\hat B_\mu$ are the gauge fields, written as
matrices:
\bea
\hat W_\mu&=&\frac{1}{2i}\vec W_\mu\cdot\vec\tau\nn\\
\hat B_\mu&=&\frac{1}{2i} B_\mu\tau^3
\eea
The corresponding field strengths are given by:
\bea
\hat W_{\mu\nu}&=&\dmu\hat W_\nu-\dnu\hat W_\mu-g[\hat W_\mu,\hat W_\nu]\nn\\
\hat B_{\mu\nu}&=&\dmu\hat B_\nu-\dnu\hat B_\mu
\eea
Their transformation laws are the following:
\bea
\hat W'_\mu&=&g_L\hat W_\mu g_L^\dagger-\frac{1}{g}g_L\dmu g_L^\dagger\nn\\
\hat B'_\mu&=&\hat B_\mu-\frac{1}{g'}g_R\dmu g_R^\dagger
\eea
and
\bea
\hat W'_{\mu\nu}&=&g_L\hat W_{\mu\nu}g_L^\dagger\nn\\
\hat B'_{\mu\nu}&=&\hat B_{\mu\nu}
\eea
The lagrangian for the bosonic fields is given by:
\be
\LL_B=\frac{1}{2}tr(\hat W_{\mu\nu}\hat W^{\mu\nu}+
\hat B_{\mu\nu}\hat B^{\mu\nu})+
\frac{v^2}{4}tr(D_\mu U^\dagger D^\mu U)
\ee
{}From the last term in the previous equation, by going to the unitary
gauge $U={\bf 1}$, we can read the mass term for the gauge vector bosons.
One introduces the
combinations:
\bea
{W^\pm_\mu}&=&\frac{(W^1_\mu\mp iW^2_\mu)}{\sqrt{2}}\nn\\
{Z_\mu}&=&\cos\theta ~W^3_\mu-\sin\theta ~B_\mu\nn\\
{A_\mu}&=&\sin\theta ~W^3_\mu+\cos\theta ~B_\mu
\eea
where the angle $\theta$ is defined by:
\be
\tan\theta=\frac{g'}{g}
\ee
The mass spectrum in the gauge vector boson sector is given in table I.

\begin{center}
\begin{tabular}{|c|c|}
\hline
&$(mass)^2$\\
\hline
$W^\pm$&$\dd\frac{v^2g^2}{4}$\\
\hline
$Z$&$\dd\frac{v^2(g^2+g'^2)}{4}$\\
\hline
$A$&0\\
\hline
\end{tabular}
\end{center}

\begin{quotation}
{\bf Table I}: electroweak gauge vector boson masses.
\end{quotation}

Now we consider the lagrangian $\LL_\psi$ for the fermionic fields,
whose quantum numbers are listed in table II.
\begin{center}
\begin{tabular}{|c|c|c|c|}
\hline
$\psi$ & $SU(2)_L$ & $Y$ & $\dd\frac{(B-L)}{2}$\\
\hline
$q_L={\left(
         \begin{array}{c}
         u_L\\
         d_L
         \end{array}
         \right)}$ &
{\underbar{2}} & 1/6 & 1/6\\
\hline
$q_R={\left(
         \begin{array}{c}
         u_R\\
         d_R
         \end{array}
         \right)}$ &

       ${\begin{array}{c}
         \underbar{1}\\
         \underbar{1}
         \end{array}
         }$ &
${
         \begin{array}{c}
         2/3\\
         -1/3
         \end{array}
         }$ &
${
         \begin{array}{c}
         1/6\\
         1/6
         \end{array}
         }$ \\
\hline
$l_L={\left(
         \begin{array}{c}
         \nu_L\\
         e_L
         \end{array}
         \right)}$ &
{\underbar{2}} & -1/2 & -1/2\\
\hline
$l_R={\left(
         \begin{array}{c}
         0\\
         e_R
         \end{array}
         \right)}$ &
${
         \begin{array}{c}
         \\
         \underbar{1}
         \end{array}
         }$ &
${
         \begin{array}{c}
         \\
         -1
         \end{array}
         }$ &
${
         \begin{array}{c}
         \\
         -1/2
         \end{array}
         }$ \\
\hline
\end{tabular}
\end{center}

\begin{quotation}
{\bf Table II}: fermions and their electroweak quantum numbers.
\end{quotation}

The kinetic terms for the fermionic fields and their minimal coupling
to the gauge vector bosons are given by:
\bea
\LL_\psi&=&i\bar q_L\gamma^\mu D_\mu q_L +
         i\bar q_R\gamma^\mu D_\mu q_R +\nn\\
        &+&i\bar l_L\gamma^\mu D_\mu l_L +
         i\bar l_R\gamma^\mu D_\mu l_R
\eea
Indices in the generation space are understood here.
The covariant derivatives acting on the left and right-handed fermions
$\psi_{L,R}$, ($\psi=q,l$) are defined below:
\bea
D_\mu\psi_L&=&(\dmu- g\hat W_\mu - g'\hat B^{(L)}_\mu)\psi_L\\
D_\mu\psi_R&=&(\dmu- g'\hat B^{(R)}_\mu)\psi_R
\eea
The combinations $\hat B^{(L,R)}_\mu$ are given by:
\bea
\hat B^{(L)}_\mu&=&\frac{1}{2i}(B-L) B_\mu\\
\hat B^{(R)}_\mu&=&\frac{1}{2i}(\tau^3+(B-L)) B_\mu
\eea

As a function of the mass eigenstates for the gauge vector bosons
(eq. (3.20)),
the lagrangian given in eq. (3.22) contains the following interaction
terms:
\be
-\frac{g}{\sqrt{2}}(W^+_\mu {J^+}^\mu+ W^-_\mu {J^-}^\mu)
-e A_\mu J^\mu_{em}
-\frac{g}{\cos\theta}Z_\mu {J_Z}^\mu
\ee
where
\bea
{J^\pm}^\mu&=&\sum_\psi \bar\psi_L\gamma^\mu\tau^\pm\psi_L\\
{J_{em}}^\mu&=&\sum_\psi Q_{em}^\psi \bar\psi\gamma^\mu\psi\\
{J_Z}^\mu&=&(J_{3L}^\mu-\sin^2\theta J_{em}^\mu)\\
{J_{3L}}^\mu&=&\sum_\psi\bar\psi_L\gamma^\mu\frac{\tau^3}{2}\psi_L
\eea
As usual, one has $\tau^{\pm}=(\tau^1\pm i\tau^2)/2$, $Q_{em}=T^3+Y$ and
$e=g\sin\theta$.

The mass terms for quarks and leptons are introduced by means
of gauge invariant
Yukawa interaction terms, $\LL_Y$. We have:
\be
\LL_Y=\bar q_L Um_q q_R + \bar l_L U m_l l_R + h.c.
\ee
$\LL_Y$ is invariant under $SU(2)_L\otimes U(1)_Y$ provided the mass
matrices $m_q$ and $m_l$ are linear combinations of ${\bold 1}$
and $\tau^3$, in the corresponding two-dimensional flavour space.
This means that one has independent
mass matrices for each separate charge $u,d,e$.

By ignoring the Higgs degree of freedom, the total lagrangian is
given by:
\be
\LL=\LL_B+\LL_\psi+\LL_Y
\ee

This model can be easily extended to account for the presence of the
Higgs particle, here identified with a real scalar field $\sigma$,
invariant under the whole gauge group $SU(2)_L\otimes U(1)_Y$ (see
eq. (3.11)).
One has simply to add to the lagrangian $\LL$, additional terms of
the form:
\bea
\LL_\sigma&=&\frac{1}{2}\dmu\sigma\dmua\sigma-V(\sigma)+\nn\\
           &+&\left[a\left(\frac{\sigma}{v}\right)
             \frac{v^2}{4}tr(D_\mu U^\dagger D^\mu U)+
             b\left(\frac{\sigma}{v}\right)^2
             \frac{v^2}{4}tr(D_\mu U^\dagger D^\mu U)+...\right]\nn\\
           &+&\left[\left(\frac{\sigma}{v}\right)\bar q_L U M_q q_R+
             \left(\frac{\sigma}{v}\right)\bar l_L U M_l l_R + h.c.\right]
\eea
Since $\sigma$ is a singlet under the entire gauge group, other
interactions could be easily added to the lagrangian $\LL_\sigma$, which
contains
just few possible terms. Notice that $a,b,...$ are
arbitrary real parameters, and $M_{q,l}$ are matrices, linear combinations
of ${\bf 1}$ and $\tau^3$, not necessarily equal to the matrices
$m_{q,l}$. $V(\sigma)$ denotes here the scalar potential for the
Higgs field.
\vskip .5truecm

The usual standard model of the electroweak interactions is a specialization
of the lagrangian $\LL+\LL_\sigma$ to the following choice of $a$, $b$,
$M_q$, $M_l$ and $V$:
\bea
a&=&2\nn\\
b&=&1\nn\\
M_q&=&m_q\nn\\
M_l&=&m_l\nn\\
V((\sigma+v)^2)&=&\frac{m_\sigma^2}{8v^2}((\sigma+v)^2-v^2)^2
\eea
With the above choice, we can easily rewrite the lagrangian
$\LL+\LL_\sigma$ in the form:
\bea
\LL_{SM}&=&\frac{1}{2}tr(\hat W_{\mu\nu}\hat W^{\mu\nu}+
\hat B_{\mu\nu}\hat B^{\mu\nu})+
\frac{1}{4}tr(D_\mu M^\dagger D^\mu M)-V(M^\dagger M)+\nn\\
&+&\LL_\psi+
\left[\bar q_L M\frac{m_q}{v} q_R + \bar l_L M \frac{m_l}{v} l_R + h.c.\right]
\eea
where:
\be
M=\sigma U=\sigma e^{\dd i\frac{\vec\xi\cdot\vec\tau}{v}}
\ee
Finally, to obtain the lagrangian for the standard model in a more familiar
form, we can perform the following field redefinition:
\be
M=
\sqrt{2}\left(\begin{array}{cc}
\varphi^0&-\varphi^+\\
\varphi^-&(\varphi^0)^*
\end{array}\right)
\ee
Taking into account the vacuum expectation value $v$ of the field $\sigma$
in eq. (3.37), we realize that
this field transformation is an allowed one and the S-matrix elements
of the theory remain unchanged. We recognize in
\be
\left(\begin{array}{c}\varphi^0\\ \varphi^-\end{array}\right)
\ee
the usual doublet of scalar fields.

Up to now, Higgs particles have not been detected, and on the
standard Higgs mass, the LEP collaborations have put a lower
bound of $60~GeV$ \cite{Higgs}.
{}From this point of view, there is no reason to prefer, as an effective
model for low-energy electroweak interactions, one particular
model among $\LL$, $\LL+\LL_\sigma$
\footnote{Bounds on the matrices $M_{q,l}$ in $\LL_\sigma$ come
from data on flavour changing processes}
, $\LL_{SM}$.
However, an important property is
enjoyed only by $\LL_{SM}$:
the renormalizability. With the increasing accumulation of
precision tests of the electroweak theory,
the compelling need to take into account quantum effects in comparing
predictions and measurements makes mandatory the use
of a consistent framework for the evaluation of the radiative corrections,
that is a renormalizable theory.
Nevertheless, as shown in the following sections, there is an interesting
use one can make of non-renormalizable, effective lagrangians, in
connection with the existence of something beyond the standard model.
\vskip .5truecm

We conclude this section with a remark. The lagrangian $\LL+\LL_\sigma$
is not the most general lagrangian invariant under $SU(2)_L\otimes U(1)_Y$,
containing up to two derivatives. By taking advantage of the formalism
of non-linear realizations, one could introduce interactions between
fermions and gauge bosons which explicitly violate the minimal form
given in eq. (3.22) \cite{Peccei}. As an example, we consider a left-handed
quark $b_L$ of electric charge $-1/3$, embedded in a $SU(2)_L$
doublet with hypercharge $Y=1/6$:
\be
\left(\begin{array}{c}0\\ b_L\end{array}\right)
\ee
The corresponding linear multiplet, defined according the eq. (2.27),
is given by:
\be
\Psi=U\left(\begin{array}{c}0\\ b_L\end{array}\right)
\ee
It is then immediate to verify that the interaction term
\be
tr(\tau^3 U^\dagger D_\mu U)\bar{b_L} \gamma^\mu b_L
\ee
is gauge invariant and provides a modification of the tree-level
standard model $Zb\bar{b}$ coupling. Non-minimal terms of this kind arise for
instance
in the low energy limit of the standard model for a large top mass [17].

\newpage

\resection{Precision Measurements and Electroweak Radiative corrections}

At the moment there is a quite astonishing agreement between the
predictions of the standard model and the whole set of data from
precision measurements. The LEP data have certainly played a major
role in testing with great accuracy the standard model theory.
The agreement has been pushed to the level of checking
the radiative corrections,
namely the predictions of the model including the most relevant quantum
effects.

Associated to the problem of computing the radiative corrections,
there is an obstacle given by the presence of infinities in
the intermediate steps of the computation. Such infinities are
dealt with by a renormalization procedure. In practice, starting from
a lagrangian $\LL(g)$ depending on a set of coupling constants $g$,
one introduces
a regulator $\Lambda$ to give a mathematical meaning to the expressions
obtained. The physical amplitudes computed in the regulated theory
$\LL(g,\Lambda)$, diverge in the limit of infinite $\Lambda$.  Order by
order in perturbation theory, a  suitable
set of counterterms $\sum_i \delta g_iC_i$ is added to the original
lagrangian $\LL(g,\Lambda)$,
so that each amplitude is finite, at that given order.
The ambiguities related to the introduction of divergent counterterms
are removed by the requirement of specific renormalization conditions,
defining the renormalized parameters of the theory.

The basic property of renormalizable theories is that the counterterms can
be absorbed by redefining the parameters of the original theory:
\be
g\to g_0=g+\delta g
\ee
In this way, at all orders in perturbation theory,
the predictions of the theory depend on a given number of
parameters, which can be determined by a finite number of
independent measurements.
In theories characterized by global or local symmetries,
a great simplification of the renormalization procedure is obtained
by adopting a regularization preserving the symmetries.
However, there are cases in which classical symmetries are
violated at the quantum level \cite{Anomaly}.

In the following, we will assume that we have computed, for a particular model,
some radiative corrections and we will discuss how the physical
quantities are affected by them.
To simplify the analysis, we will make the following
hypotheses \cite{Peskin,Terning,Golden,AB,ABJ}:

(i) The radiative effects are related to a mass
scale $M$ much greater than the electroweak scale $M_Z$.

(ii) The radiative effects are dominated by the gauge bosons self-energy
corrections, at least for a suitable set of measurable quantities
\footnote{In the following we will be mostly concerned with the
LEP I physics.}.

These assumptions are both violated in the standard model \cite{Yellow}.
However, they can be fulfilled in some of its extensions,
at least for that part of quantum effects having a non-standard origin.
Later on we will provide some examples.

We will denote by $-i\Pi_{ij}^{\mu\nu}(p)$ the set of self-energy
corrections for the gauge boson fields, evaluated at some loop order
\footnote{We are following here the presentation given in ref. \cite{AB}}.
The indices $i,j$ can take the values $0$ (for the field $B$) and
$1,2,3$ (for the fields $W^i$),
or, alternatively, the values $\gamma,~Z,~W$. One has:
\be
-i\Pi_{ij}^{\mu\nu}(p)=-i\left[\Pi_{ij}(p^2)g^{\mu\nu}+
\left(p^\mu p^\nu~~ terms\right)\right]
\ee
The terms proportional to $p^\mu p^\nu$ have no practical effect
for the LEP I physics, and they will be disregarded from now on.
The scalar function $\Pi_{ij}(p^2)$ can be expanded around
the point $p^2=0$:
\be
\Pi_{ij}(p^2)=A_{ij}+p^2F_{ij}+...
\ee
According to our assumption (i), this expansion, meaningful for
$p^2\ll M^2$, will contain real coefficients $A_{ij}$, $F_{ij}$, etc.
Moreover, since $\Pi_{ij}(p^2)$ has dimension two in units of mass,
it will be reasonable to neglect the dots in eq. (4.3), representing
terms suppressed by
positive powers of $(p^2/M^2)$.

As a consequence of the exact electromagnetism gauge invariance,
we have $A_{\gamma\gamma}=A_{\gamma Z}=0$. Then
we are left with the six independent coefficients $A_{ZZ}$, $A_{WW}$,
$F_{\gamma\gamma}$, $F_{\gamma Z}$, $F_{ZZ}$, $F_{WW}$.
The measurable quantities will depend on these six parameters.
However, three combinations of them are related to very special
observables, which, in the electroweak theory, play the role of fundamental
constants. These are given by the electromagnetic fine structure
constant $\alpha$, the Fermi constant $G_F$ and the mass of the
$Z$ gauge vector boson $M_Z$.
It is immediate to verify that the shifts in the fundamental
constants, due to the quantum corrections induced by the gauge
vector boson self-energies, are given by:
\bea
\frac{\delta\alpha}{\alpha}&=&-F_{\gamma\gamma}\\
\frac{\delta G_F}{G_F}&=&\frac{A_{WW}}{M_W^2}\\
\frac{\delta M_Z^2}{M_Z^2}&=&-\left(\frac{A_{ZZ}}{M_Z^2}+F_{ZZ}\right)
\eea
For future reference, we also give the shift for the $W$ vector boson
mass:
\be
\frac{\delta M_W^2}{M_W^2}=-\left(\frac{A_{WW}}{M_W^2}+F_{WW}\right)
\ee

\noindent
$\bullet$
{\bf Exercise}: derive eqs. (4.4-4.7).

We conclude that, in our approximation,
the parameters counting independent measurable effects,
induced at the quantum level by the underlying theory, are three combinations
among the six coefficients $A_{ZZ},~A_{WW},
{}~F_{\gamma\gamma},~F_{\gamma Z},~F_{ZZ},~F_{WW}$.

To identify these combinations, we will now compute the radiative corrections
for the three following observables: the ratio of the
gauge boson masses $M_W/M_Z$, the forward-backward
asymmetry $A_{FB}^\mu$ in $e^+e^-\to
\mu^+\mu^-$ at the $Z$ peak and the partial width of the $Z$ into charged
leptons, $\Gamma_l$.
\vskip .5truecm

\noindent
$\bullet$
$\dd\frac{M_W}{M_Z}$

We trade ${M_W}/{M_Z}$ for the observable $\Delta r_W$ defined as
follows:
\be
\left(\frac{M_W}{M_Z}\right)^2=\frac{1}{2}+
\sqrt{\frac{1}{4}-\frac{\mu^2}{M_Z^2 (1-\Delta r_W)}}
\ee
where:
\be
\mu^2=\frac{\pi\alpha(M_Z^2)}{\sqrt{2} G_F}=(38.454~GeV)^2
\ee
The definition given in eqs. (4.8-4.9) is suggested by the lowest order
relation:
\be
\left(\frac{m_W}{m_Z}\right)^2=\frac{1}{2}+
\sqrt{\frac{1}{4}-\frac{\mu_0^2}{m_Z^2}}
\ee
with:
\be
\mu_0^2=\frac{\pi\alpha_0}{\sqrt{2} G_F^0}
\ee
and
\bea
\alpha&=&\alpha_0 (1+\frac{\delta\alpha}{\alpha})\\
G_F&=&G_F^0 (1+\frac{\delta G_F}{G_F})\\
M_{W,Z}^2&=&m_{W,Z}^2 (1+\frac{\delta M_{W,Z}^2}{M_{W,Z}^2})
\eea
By combining eqs. (4.8-4.14), one obtains:
\be
\left(\frac{M_W}{M_Z}\right)^2=\left(\frac{m_W}{m_Z}\right)^2
\left(1-\frac{\sin^2\theta}{\cos 2\theta}
(\Delta r_W-\dmz+\da-\dg)\right)
\ee
On the other hand, one has:
\be
\left(\frac{M_W}{M_Z}\right)^2=\left(\frac{m_W}{m_Z}\right)^2
\left(1+\dmw-\dmz\right)
\ee
By comparing eqs. (4.15) and (4.16) and by using eqs. (4.4-4.7), one finds:
\bea
\Delta r_W&=&-\frac{\cos^2\theta}{\sin^2\theta}\left(\frac{A_{ZZ}}{M_Z^2}-
\frac{A_{WW}}{M_W^2}\right)+\nn\\
&+&\frac{\cos 2\theta}{\sin^2\theta}\left(F_{WW}-F_{33}\right)+\nn\\
&+&2\frac{\cos\theta}{\sin\theta} F_{30}
\eea
\vskip .5truecm

\noindent
$\bullet$
${\bf A_{FB}^\mu}$

Also in this case we proceed through a series of definitions
inspired to the lowest order relations:
\be
A_{FB}^\mu (p^2=M_Z^2)=3\left(\frac{g_V g_A}{g_V^2+g_A^2}\right)^2
\ee
\bea
g_V&=&-\frac{1}{2}+2\sin^2\hat\theta\\
g_A&=&\frac{1}{2}
\eea
\be
\sin^2\hat\theta=(1+\Delta k)\sin^2\tilde\theta
\ee
\bea
\sin^2\tilde\theta&=&\frac{1}{2}-
\sqrt{\frac{1}{4}-\frac{\mu^2}{M_Z^2}}\nn\\
&=&0.23145~~~~({\rm for}~~M_Z=91.175~GeV)
\eea
With these definitions, the knowledge of $A_{FB}^\mu$ is
equivalent to that of the parameter $\Delta k$, given in eq. (4.21).
To determine the latter, we focus on the neutral current
scattering process $e^+e^-\to\mu^+\mu^-$, with electrons and muons
in the right-handed polarization state.
The lowest order amplitude to this process is derived by the
interaction lagrangian (see eqs. (3.27-3.31)):
\be
\left(e_0A_\mu -g_0\frac{\sin^2\theta_0}{\cos\theta_0}Z_\mu \right) J_{em}^\mu
\ee
and is proportional to:
\be
i\left[\frac{e_0^2}{p^2}+\left(-g_0\frac{\sin^2\theta_0}{\cos\theta_0}\right)^2
\frac{1}{(p^2-m_Z^2)}\right]
\ee
The self-energy corrections induce the terms:
\bea
&-&ie_0^2\dd\frac{F_{\gamma\gamma}}{p^2}-\nn\\
&-&ig_0^2\dd\frac{\sin^4\theta_0}{\cos^2\theta_0}\dd\frac{1}{(p^2-m_Z^2)}
\dd\frac{A_{ZZ}+p^2F_{ZZ}}{(p^2-m_Z^2)}+\nn\\
&+&2i(e_0g_0)\dd\frac{sin^2\theta_0}{\cos\theta_0}
\dd\frac{F_{\gamma Z}}{(p^2-m_Z^2)}
\eea
The first term of eq. (4.24) combines with the first term in eq. (4.25),
giving a shift of the constant $\alpha$, as given by eq. (4.4).
The sum of the remaining terms, up to higher order corrections, is
given by:
\be
i\dd\frac{g_0^2}{\cos^2\theta_0}(1-F_{ZZ})(-\sin^2\theta_0)^2
\left(1+2\dd\frac{\cos\theta}{\sin\theta}F_{\gamma Z}\right)
\dd\frac{1}{(p^2-M_Z^2)}
\ee
The factor $(1-F_{ZZ})$ in the previous expression represents a universal
correction for the neutral current. (It is the analogue of the factor
$(1-F_{\gamma\gamma})$ for the electromagnetic current.) To see this,
one can consider the scattering process $\nu_e\nu_e\to\nu_\mu\nu_\mu$,
whose amplitude is proportional to:
\be
i\dd\frac{g_0^2}{\cos^2\theta_0}(1-F_{ZZ})
\dd\frac{1}{(p^2-M_Z^2)}
\ee
In conclusion, the self-energy corrections can be accounted for
by an effective neutral current lagrangian given by:
\be
\dd\frac{g_0}{\cos\theta_0}\sqrt{1-F_{ZZ}}
\left[J^\mu_{3L}-\sin^2\theta_0\left(1+\frac{\cos\theta}{\sin\theta}
F_{\gamma Z}\right)J^\mu_{em}\right]Z_\mu
\ee
{}From this lagrangian, one can read the
effective Weinberg angle $\sin^2\hat\theta$:
\be
\sin^2\hat\theta=\left(1+\dd\frac{\cos\theta}{\sin\theta} F_{\gamma Z}
\right) \sin^2\theta_0
\ee
On the other hand, from the definition of $\sin^2\tilde\theta$
given in eq. (4.21),
one has:
\be
\sin^2\tilde\theta=\sin^2\theta_0\left(1+\dd\frac{\cos^2\theta}{\cos 2\theta}
\left(\da-\dg-\dmz\right)\right)
\ee
By comparing eqs. (4.29) and (4.30) and by making use of eqs. (4.4-4.7), one
finds:
\bea
\Delta k=&-&\frac{\cos^2\theta}{\cos 2\theta}
\left(\dd\frac{A_{ZZ}}{M_Z^2}-\dd\frac{A_{WW}}{M_W^2}\right)\nn\\
&+&\dd\frac{1}{\cos 2 \theta}\dd\frac{\cos\theta}{\sin\theta} F_{30}
\eea
\vskip .5truecm

\noindent
$\bullet$
$\Gamma_l$

We define $\Gamma_l$ as follows:
\be
\Gamma_l=\dd\frac{G_F M_Z^3}{6\pi\sqrt{2}}(1+\Delta\rho)
\left(g_V^2+g_A^2\right)
\ee
To compute the parameter $\Delta\rho$, one can refer to
the effective neutral current lagrangian of eq. (4.28).
This contains the overall factor:
\be
\dd\frac{g_0}{\cos\theta_0}\sqrt{1-F_{ZZ}}
\ee
We relate this factor to the physical constants $G_F$ and $M_Z$.
One has:
\be
4\sqrt{2}G_FM_Z^2=\dd\frac{g_0^2}{\cos^2\theta_0}
\left(1+\dg+\dmz\right)
\ee
{}From the previous equation, and from eqs. (4.4-4.7), one obtains:
\be
\dd\frac{g_0^2}{\cos^2\theta_0}(1-F_{ZZ})=
4\sqrt{2}G_FM_Z^2
\left(1+\left(\dd\frac{A_{ZZ}}{M_Z^2}-\dd\frac{A_{WW}}{M_W^2}\right)\right)
\ee
Therefore, the parameter $\Delta\rho$ is given by:
\be
\Delta\rho=\dd\frac{A_{ZZ}}{M_Z^2}-\dd\frac{A_{WW}}{M_W^2}
\ee
\vskip 1.truecm
By looking at the expressions we have obtained for the
quantities $\Delta r_W$, $\Delta k$ and $\Delta\rho$,
we recognize that they depend on the following three
combinations of self-energy corrections:
\bea
\epsilon_1&=&\dd\frac{A_{ZZ}}{M_Z^2}-\dd\frac{A_{WW}}{M_W^2}\nn\\
\epsilon_2&=&F_{WW}-F_{33}\nn\\
\epsilon_3&=&\dd\frac{\cos\theta}{\sin\theta} F_{30}
\eea
We summarize this dependence below:
\bea
\Delta r_W&=&-\dd\frac{\cos^2\theta}{\sin^2\theta}\epsilon_1
+\dd\frac{\cos 2\theta}{\sin^2\theta}\epsilon_2
+2\epsilon_3\nn\\
\Delta k&=&-\dd\frac{\cos^2\theta}{\cos 2 \theta}\epsilon_1+
\dd\frac{1}{\cos 2\theta}\epsilon_3\nn\\
\Delta\rho&=&\epsilon_1
\eea
{}From the experimental values $M_W/M_Z=0.8807\pm0.0031$,
$A_{FB}^l=0.0174\pm0.0030$ and $\Gamma_l=83.52\pm0.33~MeV$,
one finds \cite{Altarelli}:
\bea
\epsilon_1&=&(0.15\pm0.41)\cdot10^{-2}\nn\\
\epsilon_2&=&(-0.71\pm0.83)\cdot10^{-2}\nn\\
\epsilon_3&=&(-0.02\pm0.56)\cdot10^{-2}
\eea
As expected, the physical quantities depend on the three fundamental constants
of the electroweak theory and on three additional parameters
carrying, in our approximation, all the information concerning
the quantum corrections
\footnote{To remove the assumptions (i) and (ii) made above, one may
consider the eqs. (4.38) as definitions of the parameters $\epsilon$'s, taking
advantage of the fact that the three observables involved are
experimentally clean. This is the point of view advocated by the authors
of ref. \cite{ABJ}. Then the relation between the $\epsilon$'s and the
radiative
corrections will depend on the particular model examined.}.
In the next section we will relate these parameters
to those which characterize the effective lagrangian of the electroweak
interactions up to O($p^4$).

\newpage
\resection{Electroweak Interactions: Higher Orders}

In section 3, we have shown how to build an effective lagrangian
for the electroweak interactions. The underlying fundamental theory
might considerably differ from what presently assumed. In particular,
particles much heavier than those characterizing the known low-energy
spectrum might exist. In this case, none of the low-energy models
described in
section 3 will be able to reproduce the predictions of the theory.
As we have seen in the previous section, in general, new heavy particles
will affect the
physical observables through their contribution to
radiative corrections. The low-energy models, as specified in
section 3, cannot account for these corrections and appropriate
extensions of them are required.

At the same time, there is an independent motivation to enlarge
the low energy models introduced up to now. In fact, the lagrangian $\LL$
of eq. (3.33) is naturally organized in a derivative
expansion, whose lowest order term is precisely given by eq. (3.33).
At the next order, gauge invariant structures with up to four derivatives
must be included in the effective lagrangian, and so on.
As a first step, we will introduce these additional terms.
To this purpose we introduce the combinations:
\bea
T&=&U\tau^3U^\dagger\nn\\
V_\mu&=&(D_\mu U) U^\dagger
\eea
They transform under $SU(2)_L\otimes U(1)_Y$ as follows:
\bea
T'&=&g_LTg_L^\dagger\nn\\
V'_\mu &=&g_LV_\mu g_L^\dagger
\eea
The covariant derivative acting on $V_\mu$ is given by:
\be
\DD_\mu V_\nu=\dmu V_\nu-g[\hat W_\mu,V_\nu]
\ee
A frequently occurring identity is:
\be
\DD_\mu V_\nu-\DD_\nu V_\mu=-g\hat W_{\mu\nu}+g'U\hat B_{\mu\nu}U^\dagger
+[V_\mu,V_\nu]
\ee
The algebraically independent $SU(2)_L\otimes U(1)_Y$ and CP invariants,
functions of the gauge vector bosons and the Goldstone fields,
containing up to four derivatives are listed below \cite{Appelquist,Longhi}:
\bea
\LL_0&=&\dd\frac{v^2}{4}[tr(TV_\mu)]^2\nn\\
\LL_1&=&i\dd\frac{gg'}{2}B_{\mu\nu}tr(T\hat W^{\mu\nu})\nn\\
\LL_2&=&i\dd\frac{g'}{2}B_{\mu\nu}tr(T[V^\mu,V^\nu])\nn\\
\LL_3&=&gtr(\hat W_{\mu\nu}[V^\mu,V^\nu])\nn\\
\LL_4&=&[tr(V_\mu V_\nu)]^2\nn\\
\LL_5&=&[tr(V_\mu V^\mu)]^2\nn\\
\LL_6&=&tr(V_\mu V_\nu)tr(TV^\mu)tr(TV^\nu)\nn\\
\LL_7&=&tr(V_\mu V^\mu)[tr(TV^\nu)]^2\nn\\
\LL_8&=&\dd\frac{g^2}{4}[tr(T\hat W_{\mu\nu})]^2\nn\\
\LL_9&=&\dd\frac{g}{2}tr(T\hat W_{\mu\nu})tr(T[V^\mu,V^\nu])\nn\\
\LL_{10}&=&[tr(TV_\mu)tr(TV_\nu)]^2\nn\\
\LL_{11}&=&tr((\DD_\mu V^\mu)^2)\nn\\
\LL_{12}&=&tr(T\DD_\mu \DD_\nu V^\nu)tr(TV^\mu)\nn\\
\LL_{13}&=&\dd\frac{1}{2}[tr(T \DD_\mu V_\nu)]^2\nn\\
\LL_{14}&=&ig\epsilon^{\mu\nu\rho\sigma}tr(\hat W_{\mu\nu}V_\rho)tr(TV_\sigma)
\eea
Before analyzing the physical meaning of the invariants $\LL_i~(i=0,...14)$,
we will discuss the arbitrariness in the choice of a particular
base of invariants.
The above base can be modified either by adding total derivatives
to the various terms, or by making use of the classical equations of motion
\cite{Georgi1,Gasser,Buch}.

To illustrate this last point, we consider an effective lagrangian
$\LL_{eff}$, depending on a single scalar field $\varphi$ and its
derivatives, of the following form:
\bea
\LL_{eff}&=&\LL_{cl}+\sum_i c_i\LL_i\\
\LL_{cl}&=&\dd\frac{1}{2}\dmu\varphi\dmua\varphi-\dd\frac{m^2}{2}\varphi^2
-V(\varphi)
\eea
We assume that the coefficients $c_i$ are of order $\slash h$.
What we have in mind is that the part $\sum_i c_i\LL_i$, together
with the O($\slash h$) corrections from $\LL_{cl}$, correctly
reproduce the results of the underlying fundamental theory, at one-loop level.

Suppose that the term $\LL_j$, $j$ being one particular among the
$i$ indices, has the form:
\be
\LL_j=(\partial^2+m^2)\varphi\cdot F(\varphi)
\ee
where $F(\varphi)$ is at least quadratic in $\varphi$ and/or
its derivatives.
We perform the following local transformation on the field $\varphi$:
\be
\delta\varphi=c_jF(\varphi)
\ee
Notice that this is an allowed transformation, so that the S-matrix
elements do not change.
One obtains:
\bea
\delta S_{cl}&=&\int dx \dd\frac{\delta\LL_{cl}}{\delta\varphi}
\delta\varphi\nn\\
&=&-\int dx \left[(\partial^2+m^2)\varphi+
\dd\frac{\delta V}{\delta\varphi}\right] c_j F(\varphi)\nn\\
&=&-\int dx c_j \LL_j -\int dx c_j\dd\frac{\delta V}{\delta\varphi}
F(\varphi)
\eea
On the other hand:
\be
\delta\left(\int dx \sum_ic_i\LL_i\right)=O({\slash h}^2)
\ee
In conclusion the transformed lagrangian is given by:
\be
\LL'_{eff}=\LL_{cl}+\sum_{i\ne j}c_i \LL_i
-c_j\dd\frac{\delta V}{\delta\varphi}F(\varphi)+O({\slash h}^2)
\ee
The net effect of the transformation given in eq. (5.9) is identical
to that obtained
by using the classical equations of motion. Up to higher order
corrections, $\LL_{eff}$ and $\LL'_{eff}$ will give rise to the same
on-shell amplitudes
\footnote{The result generalizes to higher orders \cite{Georgi1}.
If the effective theory contains terms up to the order $O({\slash h}^n)$,
and if the kinetic operator $(\partial^2+m^2)$ has already been eliminated
from all the terms of order $O({\slash h}^{m})~~(m<n)$, then the use of the
classical equation of motion gives rise to an equivalent effective
action.}.
In the following we will make use of this freedom in order to
isolate the physically independent effects related to the invariant
structures listed in eq. (5.5).

Coming back to the lagrangian $\LL$
of eq. (3.33), the equations of motion for the gauge fields are given by:

\bea
&\dmua B_{\mu\nu}+ig'\dd\frac{v^2}{4}tr(T V_\nu)-g' J^B_\nu=0\\
&\DD^\mu \hat W_{\mu\nu}-g\dd\frac{v^2}{4}
V_\nu -\dd\frac{g}{2i}\hat J^W_\nu=0
\eea
where:
\bea
J^B_\mu&=&J^{(B-L)}_\mu+J^{3R}_\mu\nn\\
       &=&\sum_\psi\bar\psi_L\gamma^\mu\dd\frac{(B-L)}{2}\psi_L+
          \sum_\psi\bar\psi_R\gamma^\mu\dd\frac{\tau^3+(B-L)}{2}\psi_R\\
\hat J^W_\mu&=&\sum_\psi
\left(\bar\psi_L\gamma^\mu\dd\frac{\tau^a}{2}\psi_L\right)\tau^a
\eea

\noindent
$\bullet$
{\bf Exercise}: derive eqs. (5.13-5.16).

Since $\dmu\partial_\nu B^{\mu\nu}=0$ and $\DD_\mu\DD_\nu \hat W^{\mu\nu}=0$,
from eqs. (5.13-5.14), one obtains:
\bea
\partial_\nu tr(TV^\nu)&=&-i\dd\frac{4}{v^2}\partial^\nu J^B_\nu\\
\DD^\nu V_\nu&=&i\dd\frac{2}{v^2}\DD^\nu\hat J^W_\nu
\eea

Both right-hand sides of eqs. (5.17) and (5.18) are classically given by
expressions of the kind:
\be
\sum_\psi\frac{m_\psi}{v^2} (\bar\psi \psi)
\ee
Therefore, as long as one considers light fermions, $m_\psi\ll v$,
they are practically
negligible and they will be discarded from now on.

\noindent
$\bullet$
{\bf Exercise}: by using the equations of motion for the gauge
fields $W$ and $B$,
given in eqs. (5.13-5.18), show that:
\bea
\LL_{11}&=&0\\
\LL_{12}&=&0\\
\LL_{13}&=&-\dd\frac{g'^2}{4} B_{\mu\nu} B^{\mu\nu}+\LL_1+\LL_8
-\LL_4+\LL_5-\LL_6+\LL_7
\eea

As a consequence of the relations (5.20-5.22), one has the equivalence among
the effective lagrangians:
\be
\LL_{eff}=\LL+\sum_{i=0}^{14} a_i \LL_i
\ee
and
\be
\LL'_{eff}=\hat\LL+\sum_{i=0}^{14} \hat a_i \LL_i
\ee
with:
\bea
\hat a_1&=&a_1+a_{13}\nn\\
\hat a_4&=&a_4-a_{13}\nn\\
\hat a_5&=&a_5+a_{13}\nn\\
\hat a_6&=&a_6-a_{13}\nn\\
\hat a_7&=&a_7+a_{13}\nn\\
\hat a_8&=&a_8+a_{13}\nn\\
\hat a_{11}&=&0\nn\\
\hat a_{12}&=&0\nn\\
\hat a_{13}&=&0
\eea
For the coefficients not listed above one has $\hat a_i=a_i$ and
the lagrangian $\hat\LL$ differs from $\LL$ by a wave function
renormalization for the vector boson $B$.

\newpage
\resection{Physical Meaning of $\LL_{eff}$}

We are now ready to discuss the meaning of the base $\{\LL_i\}$ given
in eq (5.5). The physical content of the base is more
transparent in the unitary gauge, $U={\bf 1}$, where all the
invariants $\LL_i$ collapse into polynomials of the gauge fields.
All the invariants contain at most quartic terms in the gauge fields,
but they can be grouped appropriately, depending on their expansion which
can start from two, three or four gauge vector bosons. The structures
containing bilinear terms in the gauge fields are just six:
\bea
a_0\LL_0&=&-\dd\frac{1}{4}a_0 v^2(g W^3_\mu-g'B_\mu)^2+...\nn\\
a_1\LL_1&=& \dd\frac{1}{2} a_1 gg'B_{\mu\nu}(\dmua {W^3}^\nu-\partial^\nu
{W^3}^\mu)+...\nn\\
a_8\LL_8&=&-\dd\frac{1}{4} a_8 g^2(\dmu {W^3}_\nu-\partial_\nu
{W^3}_\mu)^2+...\nn\\
a_{11}\LL_{11}&=&-\dd\frac{1}{2} a_{11}[g^2(\dmu {W^1}^\mu)^2+
g^2(\dmu {W^2}^\mu)^2 +(\dmu (g {W^3}^\mu-g'B^\mu))^2]+...\nn\\
a_{12}\LL_{12}&=& -a_{12}\dmu\partial_\nu (g {W^3}^\nu-g'B^\nu)\cdot
(g {W^3}^\mu-g'B^\mu)+...\nn\\
a_{13}\LL_{13}&=& -\dd\frac{1}{2} a_{13}\dmu (g {W^3}_\nu-g'B_\nu)\cdot
\dmua (g {W^3}^\nu-g'B^\nu)+...\nn\\
\eea
The dots stand for trilinear and quadrilinear terms in
the gauge vector bosons. They are there, together with the terms
containing the would-be Goldstone bosons to ensure the gauge invariance
of each structure. It is straightforward to compute the contributions of the
above terms to the various two-point functions. We obtain:
\bea
-i\Pi_{\mu\nu}^{11}&=&-i a_{11} g^2 p_\mu p_\nu\nn\\
-i\Pi_{\mu\nu}^{33}&=&-\dd\frac{i}{2} a_0 v^2 g^2 g_{\mu\nu}\nn\\
&&-i(a_8+a_{13}) g^2 (p^2 g_{\mu\nu}-p_\mu p_\nu)\nn\\
&&-i(a_{11}-2a_{12}+a_{13})g^2p_\mu p_\nu\nn\\
-i\Pi_{\mu\nu}^{30}&=&\dd\frac{i}{2} a_0 v^2 g g' g_{\mu\nu}\nn\\
&&+i(a_1+a_{13}) gg' (p^2 g_{\mu\nu}-p_\mu p_\nu)\nn\\
&&+i(a_{11}-2a_{12}+a_{13})gg'p_\mu p_\nu\nn\\
-i\Pi_{\mu\nu}^{00}&=&-\dd\frac{i}{2} a_0 v^2 {g'}^2 g_{\mu\nu}\nn\\
&&-i a_{13} {g'}^2 (p^2 g_{\mu\nu}-p_\mu p_\nu)\nn\\
&&-i(a_{11}-2a_{12}+a_{13}){g'}^2p_\mu p_\nu
\eea
One may be surprised by the fact that, apparently, beyond
the lowest order represented by the lagrangian $\LL$, the gauge-invariant
independent terms affecting the two-point functions of the theory
are six and not three as expected on the basis of our discussion
about the radiative corrections. Notice that, as already
observed, the present discussion is also based on an expansion
in powers of momenta where we are keeping exactly the same order
as in eq. (4.3). The apparent paradox is solved by recalling that,
with the use of the equations of motion, $\LL_{11}$, $\LL_{12}$ and
$\LL_{13}$ can be eliminated by suitably redefining the parameters
of the effective lagrangian, that is by using the effective
theory $\LL'_{eff}$ defined in eqs. (5.24-5.25). If we insist in using
the complete base, with non-vanishing $a_{11}$, $a_{12}$ and $a_{13}$,
things will arrange in such a way that the physical quantities
will depend only on three combinations among the six parameters
$a_{0}$, $a_{1}$, $a_{8}$, $a_{11}$, $a_{12}$ and $a_{13}$.
This is already evident from eqs. (6.2): the parameters
$a_{11}$ and $a_{12}$ always multiply the harmless terms proportional to
$p_\mu p_\nu$ and the parameter $a_{13}$ can be absorbed in a
redefinition of $\alpha$ and $M_Z$.

\noindent
$\bullet$
{\bf Exercise}: compute the contribution of $\LL_{eff}=
\LL+\sum_{i=0}^{14} a_i \LL_i$ to the parameters $\epsilon_1$,
$\epsilon_2$ and $\epsilon_3$. One finds:

\bea
\epsilon_1&=&2 a_0\nn\\
\epsilon_2&=&-g^2 (a_8+a_{13})\nn\\
\epsilon_3&=&-g^2 (a_1+a_{13})
\eea

The above relations show that the coefficients $a_0$, $(a_8+a_{13})$ and
$(a_1+a_{13})$ are directly related to the observables discussed in section 4,
and are therefore appropriate in parametrizing precision measurements
performed at LEP I.

\vskip .5truecm

The other terms of the base $\LL_i~(i=0,14)$ can be discussed
along similar lines.
The invariants whose expansion starts with three gauge fields are
$\LL_2$, $\LL_3$, $\LL_9$ and $\LL_{14}$.
Again, the use of the equations of motion can help in analyzing
the physical effects. It turns out that \cite{DeRujula}:
\bea
\LL_2&=&i g'^2 J^B_\mu tr(TV^\mu)+({\rm redefinitions~of~}
\delta Z_1,a_0,a_1)\nn\\
\LL_3&=&i g^2 tr(\hat J^W_\mu V^\mu)+({\rm redefinitions~of~}\delta v^2,
\delta Z_2,a_1)
\eea
The parameters $\delta Z_1$, $\delta Z_2$ and $\delta v^2$ correspond
to wave function renormalizations for the fields $B_\mu$, $W_\mu$ and $\xi$,
respectively.
The equations (6.4) bring in new invariant structures,
not considered up to now, containing fermionic vertices.
In this new base one would have universal corrections to the
fermionic vertices, leading however to the same physical
predictions obtained in the framework of the original,
purely bosonic base.

The parameters $a_2$, $a_3$, $a_9$ and $a_{14}$ might
be useful in discussing the LEP II physics.
Indeed, one can parametrize the most general couplings of two charged
vector bosons with a neutral vector boson according to the effective
lagrangian \cite{Hagi}:
\bea
\frac{\LL_{WWN}}{g_{WWN}}&=&i {g_1}^N ({W^{\dagger}}_{\mu\nu}
                          W^{\mu}N^{\nu}-W_{\mu\nu}
                          {W^{\dagger}}^{\mu}N^{\nu})
+i k_N {W^{\dagger}}_{\mu}W_\nu N^{\mu\nu}\nn\\
&&+i\frac{\lambda_N}{M_W^2}{W^\dagger}_{\lambda\mu} {W^\mu}_\nu
N^{\nu\lambda}+
g_5^N \epsilon^{\mu\nu\rho\sigma}({W^\dagger}_\mu {\partial}_\rho
W_\nu-\partial_\rho {W^\dagger}_\mu W_\nu) N_\sigma
\eea
Here $N_\mu$ stands for the photon $A$ or the $Z$ field,
$W_\mu$ is the $W^-$ field, $W_{\mu\nu}=\partial_\mu W_\nu-
\partial_\nu W_\mu$ and similarly for $N_{\mu\nu}$.
The four terms in the previous equation are the most general
$CP$ invariant terms one can build out of vector fields
with vanishing divergence. The first three couplings are
separately invariant under $P$ and $C$ transformations,
whereas the last one violates both $P$ and $C$. Additional
terms can be added if $CP$ violation is allowed \cite{Hagi}.
The conventional choice for the overall normalization constants
$g_{WWN}$ ($N=\gamma,~Z$) is:
\bea
g_{WW\gamma}&=&-e\nn\\
g_{WWZ}&=&-g\cos\theta
\eea
The first term in eq. (6.5) has the form of a minimal coupling
of the charged $W$ current to the photon or the $Z$ field.
The parameters $g_1^\gamma$ and $g_1^Z$ represent the electromagnetic
and "$Z$" charges of the $W$, in units of $g_{WW\gamma}$ and
$g_{WWZ}$, respectively.
The coefficients $k_N$ and $\lambda_N$ are related to the "magnetic"
moments of the $W$, an anomalous "magnetic" moment occurring
if $k_N\neq 1$ or $\lambda_N\neq 0$.

A direct access to the above parameters will be provided by
$W$ pair production in $e^+e^-$ collisions at the future LEP II
facility. The differential angular distribution of the produced
$W$'s turns out to be particularly sensitive to the chosen set
of $CP$ conserving couplings \cite{Hagi}.

By identifying the trilinear gauge boson interaction terms in the effective
lagrangian $\LL_{eff}$, eq. (5.23), we can readily express the coefficients
of $\LL_{WWN}$ in terms of the parameters $a_i$. One obtains:
\bea
g_1^\gamma-1&=&0\nn\\
g_1^Z-1&=&-\frac{g^2}{\cos^2\theta}a_3\nn\\
k_\gamma-1&=&g^2(a_2-a_3-a_9)-\epsilon_2+\epsilon_3\nn\\
k_Z-1&=&\frac{g^2}{\cos^2\theta}[\cos^2\theta(-a_3-a_9)
-\sin^2\theta~ a_2]-\epsilon_2-\tan^2\theta~ \epsilon_3\nn\\
\lambda_\gamma&=&0\nn\\
\lambda_Z&=&0\nn\\
g_5^\gamma&=&0\nn\\
g_5^Z&=&-\frac{g^2}{\cos^2\theta} a_{14}
\eea
The relation $g_1^\gamma=1$ expresses the exact conservation of the
electromagnetic charge. The coefficients  $\lambda_N$ are vanishing
since $\LL_{eff}$ contains terms up to the fourth order in momenta
or gauge fields.

The parameters $a_2$, $a_3$ and $a_9$ or any set of independent
combinations, together with the coefficient $a_{14}$, appear to play
in LEP II physics the same role covered by the $\epsilon$'s parameters
in LEP I measurements. Notice that invariance under isospin would
require $a_9=a_{14}=0$.

It may be observed that the present bounds on the $\epsilon$'s
parameters, summarized in eq. (4.39), translate into a bound
of few percents on $a_0$, $(a_8+a_{13})$ and $(a_1+a_{13})$.
On the other hand,
it seems reasonable to assume that, in any sensible theory, the parameters
$a_i$ are all of the same order of magnitude.
It would be a rather surprising result to find that, for instance,
$a_2$ and $a_3$ are larger than $a_0$, $(a_8+a_{13})$
and $(a_1+a_{13})$ by a factor
10 or more. On this basis, one would expect that the deviations
listed in eq. (6.7) might be at most of order $10^{-2}$-$10^{-1}$, and thus
probably
too small to be appreciated at LEP II \cite{DeRujula}.
Indeed, if we were to use
"naive dimensional analysis" \cite{Manohar} to estimate the
size of the effects at LEP II, by assuming a range of
validity for $\LL_{eff}$ extending up to energies close to $4\pi v$,
we would guess for the dimensionless
coefficients $a_i$ values of order one in units of $(1/16\pi^2)$,
even smaller than those indicated by the LEP I data.

It is maybe useful to recall that this kind of
considerations requires that the scale of new physics is considerably
higher than $2 M_W$ and, in any event, it is certainly not a substitute
for the experimental tests \cite{Zeppe}.
\vskip .5truecm

Finally, $\LL_4$, $\LL_5$, $\LL_6$, $\LL_7$ and $\LL_{10}$
contain only quadrilinear terms in the gauge boson fields.
We could think to fix them, at least in principle,
by means of scattering experiments among gauge vector bosons,
to occur at the future LHC/SSC facilities \cite{Valencia}.
Notice that all the invariants but $\LL_{14}$ are invariant under parity.

\newpage
\resection{The Matching Conditions}

The effective lagrangian $\LL_{eff}$ defined in the previous
section is able to reproduce
the quantum effects originating from the heavy sector of an
underlying fundamental theory of the electroweak interactions.
Such a piece of information is contained in the coefficients
$a_i$ of the invariants $\LL_i$. In particular, $a_0$, $a_1$,
$a_8$, $a_{11}$, $a_{12}$ and $a_{13}$ are related, as we have seen, to
the self-energy corrections. Corrections to the three and four point
functions may be discussed as well.

Before concluding these lectures, we will show, in an example,
how to deduce, in practice, the coefficients $a_i$ from the
knowledge of the fundamental theory.  This is done by means of the
so called matching procedure \cite{Georgi3}. It consists in equating
amplitudes evaluated in the fundamental theory and in the effective
one. At the end, the parameters of the effective theory will be run
from the scale where the matching has taken place, down to the
energy where one will use the effective lagrangian.

As a simple example, we consider a fundamental theory consisting
just of the standard model with a fourth generation of
light leptons $(M_l\le M_Z)$ and heavy quarks $(M \gg M_Z)$.
We denote by $M$ the common mass
of the quarks.
We will restrict the discussion to the two point functions
in the gauge boson sector evaluated at one-loop order,
the extension to the other Green functions
being straightforward
\footnote{The effective theory will also contain a Wess-Zumino term \cite{WZ}
whose gauge variation cancels the anomaly produced by the light leptons
\cite{D'Hoker}.}.

The matching condition is given by:

\be
{\Pi_C}^{ij}_{\mu\nu}+{\Pi_L}^{ij}_{\mu\nu}=
{\Pi_{\hat C}}^{ij}_{\mu\nu}+{\Pi_a}^{ij}_{\mu\nu}
\ee

Here the left-hand side refers to the fundamental
theory and the right-hand side refers to the effective one.
${\Pi_C}$ is the contribution of the counterterms
of the fundamental theory, derived from:
\bea
&-\dd\frac{1}{4}(1+\delta Z_1) B_{\mu\nu}B^{\mu\nu}
-\dd\frac{1}{4}(1+\delta Z_2) W^i_{\mu\nu}{W^i}^{\mu\nu}+\nn\\
&+\dd\frac{1}{8}(v^2+\delta v^2)g^2(W^1_\mu{W^1}^\mu+
W^2_\mu{W^2}^\mu)+
\dd\frac{1}{8}(v^2+\delta v^2)(gW^3_\mu-g'B_\mu)^2
\eea
${\Pi_L}$ is the loop contribution from the heavy quark doublet.
${\Pi_a}$ is the contribution from the invariants $\LL_i$,
already evaluated in eq. (6.2). Finally, ${\Pi_{\hat C}}$ is the
contribution from possible finite counterterms in the effective
theory, of exactly the same form as those given in eq. (7.2),
with renormalization constants $\delta\hat{Z_1}$, $\delta\hat{Z_2}$,
$\delta\hat{v^2}$. Such terms may be present for a particular choice
of renormalization conditions.
Two comments are in order.

We have not included the loop contributions from the light sector.
These are exactly the same for the fundamental and the effective theories,
and can be simply dropped out from both sides of eq. (7.1).
Actually, this is the reason why at low energy we can define
an effective lagrangian. The amplitude computed in the fundamental
theory in the large mass limit and the amplitude
evaluated with the lagrangian obtained
by simply suppressing the heavy fields, have equal absorptive
parts in the various channels.
Their difference is an analytic function, which, for momenta
much lower than the scale $M$, can be approximated with a polynomial.

In principle, in a generic renormalization scheme, additional
finite counterterms have to be added to the loop contribution
from the fundamental theory, to properly ensure the validity
of the Ward identities. If the regularization procedure
does not respect the gauge symmetry,
the Ward identities will be apparently
broken by a loop computation and they will have to be repaired
by properly adding finite terms. In the following, we will use
dimensional regularization with $\{\gamma_\mu,\gamma_5\}=0$ in all dimensions,
which automatically accounts for the non-anomalous Ward identities and which,
in the example given, can be safely applied.

The contributions from the counterterms  of eq. (7.2) are given by:
\bea
-i{\Pi_C}_{\mu\nu}^{11}=-i{\Pi_C}_{\mu\nu}^{22}=-i{\Pi_C}_{\mu\nu}^{33}&=&
i\dd\frac{\delta v^2}{4} g^2 g_{\mu\nu}
-i\delta Z_2 (p^2 g_{\mu\nu}-p_\mu p_\nu)\nn\\
-i{\Pi_C}_{\mu\nu}^{30}&=&-i\dd\frac{\delta v^2}{4} gg' g_{\mu\nu}\nn\\
-i{\Pi_C}_{\mu\nu}^{00}&=&i\dd\frac{\delta v^2}{4}{g'}^2 g_{\mu\nu}
-i\delta Z_1 (p^2 g_{\mu\nu}-p_\mu p_\nu)\nn\\
\eea
Those from the counterterms of the effective theory are immediately
derived from the above equations.
Finally, the one-loop contribution from the heavy quark doublet
is given by:
\bea
-i{\Pi_L}_{\mu\nu}^{11}=-i{\Pi_L}_{\mu\nu}^{22}=-i{\Pi_L}_{\mu\nu}^{33}&=&
\dd\frac{-i}{16\pi^2}g^2
\biggl\{3M^2\left(A+ln\dd\frac{M^2}{\mu^2}\right) g_{\mu\nu}
\nn\\
&&+\left[-\left(A+ln{\dd\frac{M^2}{\mu^2}}\right)
-{\dd\frac{1}{2}}\right](p^2 g_{\mu\nu}-p_\mu p_\nu)
\nn\\
&&-{\dd\frac{1}{2}}p_\mu p_\nu\biggr\}\\
-i{\Pi_L}_{\mu\nu}^{30}&=&
\dd\frac{-i}{16\pi^2}g g'
\biggl\{-3M^2\left(A+ln\dd\frac{M^2}{\mu^2}\right) g_{\mu\nu}
\nn\\
&&+{\dd\frac{1}{2}}(p^2 g_{\mu\nu}-p_\mu p_\nu)
\nn\\
&&+{\dd\frac{1}{2}}p_\mu p_\nu\biggr\}\\
-i{\Pi_L}_{\mu\nu}^{00}&=&
\dd\frac{-i}{16\pi^2}g'^2
\biggl\{3M^2\left(A+ln\dd\frac{M^2}{\mu^2}\right) g_{\mu\nu}
\nn\\
&&+\left[-\dd\frac{11}{9}\left(A+ln{\dd\frac{M^2}{\mu^2}}\right)
-{\dd\frac{1}{2}}\right](p^2 g_{\mu\nu}-p_\mu p_\nu)
\nn\\
&&-{\dd\frac{1}{2}}p_\mu p_\nu\biggr\}
\eea
where
\bea
A&=&-\dd\frac{2}{4-d}+\gamma_E-ln 4\pi\nn\\
\gamma_E&\simeq&0.577
\eea
and $\mu$ denotes the scale parameter of dimensional regularization.
The expressions given above have been obtained in the limit
$p^2\ll M^2$, neglecting positive powers of $p^2/M^2$.
We are now ready to solve the matching conditions. From eq. (7.1), by
using eqs. (7.3-7.6) and (6.2), one obtains:
\bea
\delta v^2-\delta {\hat v}^2&=&\frac{1}{16\pi^2}12 M^2
\left(A+ln\frac{M^2}{\mu^2}\right)\nn\\
\delta Z_2-\delta {\hat Z}_2&=&\frac{g^2}{16\pi^2}
\left(A+ln\frac{M^2}{\mu^2}+\frac{1}{2}\right)\nn\\
\delta Z_1-\delta {\hat Z}_1&=& g'^2 a_{13}+\frac{g'^2}{16\pi^2}
\left[\frac{11}{9}\left(A+ln\frac{M^2}{\mu^2}\right)+\frac{1}{2}\right]
\eea
and
\bea
a_0&=&0\nn\\
a_1+a_{13}&=&\dd\frac{1}{16\pi^2}\left(-\dd\frac{1}{2}\right)\nn\\
a_8+a_{13}&=&0\nn\\
a_{11}&=&\dd\frac{1}{16\pi^2}\left(-\dd\frac{1}{2}\right)\nn\\
a_{13}&=&2a_{12}
\eea
As expected, these relations are already sufficient to determine
the contribution of the heavy quark doublet to the
parameters $\epsilon$'s. One finds \cite{Bertolini}:
\bea
\delta\epsilon_1&=&0\nn\\
\delta\epsilon_2&=&0\nn\\
\delta\epsilon_3&=&+\dd\frac{g^2}{32\pi^2}
\eea
Notice that, since the parameters $\epsilon_1$ and $\epsilon_2$
are associated to isospin breaking effects (see eqs. (4.37)),
they receive a vanishing contribution from a degenerate quark doublet.

\noindent
$\bullet$
{\bf Exercise}: compute the contribution to the $\epsilon$'s parameters
from a degenerate doublet of heavy leptons. Do the overall contribution
to $\epsilon_3$ due to a whole fourth generation of heavy degenerate
fermions vanish?

The case analyzed above provides an example of non-decoupling.
If decoupling applies, the effects of a large mass limit
are just the appearance of higher dimensional operators, with
coefficients suppressed by inverse powers of the large mass, and
a renormalization of the parameters \cite{Carazzone}.
In this case the physics associated with the heavy scale
decouples from the low energy theory. Instead, the effects
described in eqs. (7.9)-(7.10) are neither suppressed nor absorbable in
a redefinition of the fundamental constants. The point is that,
in order to have decoupling, one is not allowed to let a
dimensionless coupling grow with the heavy mass. On the other
hand, this is just our case, since, in order to preserve
the gauge invariance, the large mass limit for the quark doublet
has been implicitly achieved by increasing the corresponding Yukawa
coupling.
\footnote{
The effective lagrangian we have dealt with so far has been constructed
by applying the formalism of non-linear realizations sketched in
section 2. According to a widely accepted point of view, going
beyond the SM, one should use non-linear realizations whenever
the decoupling theorem does not apply, the opposite case
requiring the use of linear realizations. Although this choice may be
in practice convenient, it is not so compelling.
In fact, the use of non-linear realizations is requested when
the degrees of freedom of the low energy effective theory
cannot be assembled into linear multiplets of the symmetry group $G$.
When this happens the low energy theory is non-renormalizable,
which is indeed the case if the decoupling theorem does not apply.
On the other hand, when the light degrees of freedom can be arranged
into linear multiplets of $G$, which may happen whether or not the
decoupling theorem applies, one can choose to work with linear realizations
or non-linear ones, the two being related by an allowed field transformation.
For instance, the case of a vector-like doublet of heavy quarks
can still be discussed in the framework of the effective lagrangian
of eq. (5.6).
In this case, the effects of the doublet decouple and
the coefficients $a_i$ will contain inverse power of the heavy scale
\cite{Bagger}.}

By imposing the matching conditions on a sufficient number of gauge
bosons Green functions, one can determine the whole set of
parameters $a_i~(i=0,...14)$ and we have collected the results
in table III.

\begin{center}
\begin{tabular}{|c|c|c|c|}
\hline
& LARGE $m_\sigma$\cite{Longhi} & LARGE $m_{t'}=m_{b'}=M$\cite{D'Hoker}
& LARGE $m_t\gg m_b$\cite{me}\\
& $L=ln(m_\sigma/\mu)$ & & $L=ln(m_t/\mu)$\\
\hline
$a_0$&$-\dd\frac{3}{4}g'^2 L $&$ 0$ &$\dd\frac{3}{2}\frac{m_t^2}{v^2}$\\
\hline
$a_1$&$-\dd\frac{1}{6} L$&$-\dd\frac{1}{2} $&$+\dd\frac{1}{3}L-\frac{1}{4}$\\
\hline
$a_2$&$-\dd\frac{1}{12} L $&$-\dd\frac{1}{2} $&$+\dd\frac{1}{3}L-
\dd\frac{3}{4}$\\
\hline
$a_3$&$+\dd\frac{1}{12} L $&$+\dd\frac{1}{2} $&$+\dd\frac{3}{8}$\\
\hline
$a_4$&$+\dd\frac{1}{6} L $&$+\dd\frac{1}{4} $&$+L-\dd\frac{5}{6}$\\
\hline
$a_5$&$+\dd\frac{1}{12} L $&$-\dd\frac{1}{8} $&$-L+\dd\frac{23}{24}$\\
\hline
$a_6$&$ 0 $&$0 $&$-L+\dd\frac{23}{24}$\\
\hline
$a_7$&$ 0 $&$0 $&$+L-\dd\frac{23}{24}$\\
\hline
$a_8$&$ 0 $&$0 $&$+L-\dd\frac{7}{12}$\\
\hline
$a_9$&$ 0 $&$0 $&$+L-\dd\frac{23}{24}$\\
\hline
$a_{10}$&$ 0 $&$0 $&$-\dd\frac{1}{64}$\\
\hline
$a_{11}$&$ 0 $&$-\dd\frac{1}{2}  $&$-\dd\frac{1}{2}$\\
\hline
$a_{12}$&$ 0 $&$0 $&$-\dd\frac{1}{8}$\\
\hline
$a_{13}$&$ 0 $&$0 $&$-\dd\frac{1}{4}$\\
\hline
$a_{14}$&$ 0 $&$0 $&$+\dd\frac{3}{8}$\\
\hline
\end{tabular}
\end{center}

\begin{quotation}
{\bf Table III}: $a_i$ coefficients - in units of ($1/16\pi^2$) -  for
three particular limits of the SM. The scale $\mu$ is some intermediate scale
between the low external momenta and the large mass.
\end{quotation}
\vskip .7truecm

As a last step,
to fully define the low energy effective lagrangian, one has to specify
the renormalization group equations (RGE) according to which
the various parameters run from energy scales close to $M$ down
to lower energies. By working in the effective theory and by
choosing vanishing renormalization constants $\delta {\hat Z}_1$,
$\delta {\hat Z}_2$ and $\delta {\hat v}^2$, one can readily conclude
that no contribution from the heavy quark doublet survives in the RGE's,
at one-loop order. The one-loop RGE's are then determined by the contribution
of the light sector. We find instructive to recover this conclusion by
working at the level of the fundamental theory, focusing
just on the possible contribution from the heavy particles.
As an example, we will compute such a contribution for the
$g$ gauge coupling constant, for the previously analyzed case of
a new doublet of heavy quarks. As far as the fermionic contribution
is concerned, the relation between the unrenormalized coupling $g_0$
and the renormalized one $g$ is given by:
\be
g_0=\mu^{\frac{4-d}{2}}\left(\frac{g}{\sqrt{Z_2}}\right)
\ee
where $Z_2=1+\delta Z_2$ and, since we are interested in the running of
$g$ due to the heavy quark doublet, $\delta Z_2$ is the renormalization
constant given in eq. (7.8). Actually, eq. (7.8) gives the
combination $\delta Z_2-\delta {\hat Z}_2$ and, to proceed, we have to
specify the finite counterterm $\delta {\hat Z_2}$, that is the
renormalization scheme we are going to adopt.
For instance, in the $\overline{\rm MS}$ scheme, we would choose $\delta
{\hat Z_2}$
in such a way to obtain $\delta Z_2=(g^2/16\pi^2) A$, a mass independent
renormalization constant. Here we prefer to make the unusual but
simpler choice $\delta {\hat Z_2}=0$, to obtain:
\be
\delta Z_2=\frac{g^2}{16\pi^2}
\left(A+ln\frac{M^2}{\mu^2}+\frac{1}{2}\right)
\ee
To compute the contribution of the heavy quark doublet to the $\beta$
function of $g$, we act on both sides of eq. (7.11) with $\mu d/d\mu$.
We find:
\def\gz{\dd\left(\frac{g}{\sqrt{Z_2}}\right)}
\def\dg{\dd\frac{\partial}{\partial g}}
\def\dM{\dd\frac{\partial}{\partial M}}
\def\mdem{\mu\dd\frac{\partial}{\partial\mu}}
\def\mdm{\mu\dd\frac{d}{d\mu}}
\be
0=\frac{4-d}{2}\gz+\dg\gz\mdm g+\mdem\gz+\dM\gz\mdm M
\ee
By explicitly evaluating the right-hand side of the previous equation
and by taking the limit $d\to 4$, we obtain:
\bea
\beta (g) &=&\mdm g\nn\\
&=& \frac{g^3}{16\pi^2}\frac{1}{M} \left(\mdm M\right) +O(g^5)
\eea
Eq. (7.14) shows that
the usual contribution from the quark doublet, $g^3/16\pi^2$,
drops from the final result, which is of higher order in $g$.
More precisely, this cancellation is due to the term proportional
to $\mdem\gz$ in eq. (7.13). Usually, in a mass independent
renormalization scheme, this term and the last one of eq. (7.13),
are absent. On the other hand,
the renormalization scheme chosen here is particularly
suitable to discuss the effect of the heavy doublet, since in this
case the effect simply disappears (at the order $g^3$), as expected.

Indeed, nothing prevents the use of a mass independent
scheme. For instance, in the $\overline{\rm MS}$ scheme one has
$\delta Z_2=(g^2/16\pi^2) A$ and we would have obtain:
\be
\beta (g) =\frac{g^3}{16\pi^2}~~~,
\ee
the usual result for a quark doublet. However, for the matching
conditions to hold, this would have required the presence of the finite
counterterm:
\be
\delta {\hat Z}_2=-\frac{g^2}{16\pi^2}\left(ln\frac{M^2}{\mu^2}+
\frac{1}{2}\right)
\ee
In turn, this finite counterterm
could have been absorbed in the effective coupling:
\be
g^*=g\left(1+\frac{g^2}{32\pi^2}\left(ln\frac{M^2}{\mu^2}+
\frac{1}{2}\right)\right)
\ee
As one may easily verify, $g^*$ runs according to the $\beta$
function defined in eq. (7.14), not containing any $g^3$ term.

Other cases can be studied along
similar lines. The invariant structures $\LL_i$ also occur
in the low energy effective lagrangian derived from the standard electroweak
theory in the limit of an heavy Higgs \cite{Appelquist,Longhi} and in
the limit of an heavy top quark \cite{me} and we have listed
the corresponding results in table III. In principle, any
extension of the standard model, characterized by heavy particles,
will leave its peculiar mark at low energies through a specific
contribution to the parameters $a_i$.
\vskip 2truecm

\noindent
{\bf ACKNOWLEDGMENT}:  I would like to thank R. Casalbuoni, P. Chiappetta,
A. Deandrea, S. De Curtis, N. Di Bartolomeo, D. Dominici, R. Gatto,
L. Maiani, A. Masiero for many stimulating discussions on the subject
of these lectures. A special
thank goes to the organizers of the Seminar, G. Marchesini and E. Onofri,
as well as to the secretariat, for the very nice hospitality enjoyed in Parma.


\newpage
\vspace{1cm}

\begin{thebibliography}{99}
\bibitem{Gursey}
F. Gursey, Nuovo Cim. 230 (1960) 230

\bibitem{Weinberg}
S. Weinberg, Physica 96A (1979) 327

\bibitem{Gasser}
J. Gasser and H. Leutwyler, Ann. Phys. 158 (1984) 142;
Nucl. Phys. B250 (1985) 465.

\bibitem{Leut}
H. Leutwyler, talk given at the XXVI Int. Conf. on High Energy Physics,
Dallas, Aug. 1992, preprint BUTP-92/42.

\bibitem{Ecker}
For a recent review see, for instance: G. Ecker, preprint CERN-TH.6660/92
and references therein

\bibitem{Terning}
B. Holdom and J. Terning, Phys. Lett. 247B (1990) 88

\bibitem{Golden}
M. Golden and L. Randall, Nucl. Phys. B361 (1991) 3

\bibitem{Georgi2}
H. Georgi, Nucl. Phys. B363 (1991) 301

\bibitem{Holdom}
B. Holdom, Phys. Lett. B258 (1991) 156

\bibitem{DeRujula}
A. De Rujula, M.B. Gavela, P. Hernandez and E. Masso,
preprint CERN-TH-6272-91, FTUAM-91-31.

\bibitem{Bernard}
T. Appelquist and C. Bernard, Phys. Rev. D22 (1980) 200

\bibitem{Appelquist}
A.C. Longhitano, Phys. Rev. D22 (1980) 1166;
T. Appelquist in Gauge Theories and experiments at High Energies,
ed. by K.C. Brower and D.G. Sutherland, Scottish University Summer
School in Physics, St. Andrews (1980).

\bibitem{Longhi}
A.C. Longhitano, Nucl. Phys. B188 (1981) 118

\bibitem{Veltman}
M. Veltman, Acta Phys. Pol. B8 (1977) 475;
B.W. Lee, C. Quigg and H.B. Thacker, Phys. Rev. D16 (1977) 1519;
R. Casalbuoni, D. Dominici and R. Gatto, Phys. Lett. 147B (1984) 419;
M.B. Einhorn, Nucl. Phys. B246 (1984) 75.

\bibitem{Carazzone}
T. Appelquist and J. Carazzone, Phys. Rev. D11 (1975) 2856.

\bibitem{D'Hoker}
E. D'Hoker and E. Fahri, Nucl. Phys. B248 (1984) 59;
Nucl. Phys. B248 (1984) 77.

\bibitem{Steger}
G.L. Lin, H. Steger and Y.P. Yao, Phys. Rev. D44 (1991) 2139

\bibitem{me}
F. Feruglio, L. Maiani and A. Masiero, Nucl. Phys. B387 (1992) 523

\bibitem{D'Hoker2}
E. D'Hoker, Phys. Rev. Lett. 69 (1992) 1316

\bibitem{Anomaly}
S.L. Adler, Phys. Rev. 177 (1969) 2426;
J.S. Bell and R. Jackiw, Nuovo Cim. 60A (1969) 47;
W.A. Bardeen, Phys. Rev. 184 (1969) 1848.

\bibitem{WZ}
J. Wess and B. Zumino, Phys. Lett. 37B (1971) 95;
E. Witten, Nucl. Phys. B223 (1983) 422.

\bibitem{Technicolor}
S. Weinberg, Phys. Rev. D19 1277 (1979);
L. Susskind, Phys. Rev. D20 (1979) 2619;
E. Fahri and L. Susskind, Phys. Rep. 74 (1981) 279.

\bibitem{Chadha}
S. Chadha and M. Peskin, Nucl. Phys. B185 (1981) 61;
M. Peskin, Nucl. Phys. B175 (1980) 197;
J. Preskill, Nucl. Phys. B177 (1981) 21;
R. Casalbuoni, D. Dominici, S. De Curtis, N. Di Bartolomeo, F. Feruglio
and R. Gatto, Phys. Lett. B285 (1992) 103.

\bibitem{Georgi3}
For a discussion in the framework of the electroweak interactions,
see ref. \cite{Georgi2}.

\bibitem{Manohar}
A. Manohar and H. Georgi, Nucl. Phys. B234 (1984) 189;
H. Georgi and L. Randall, Nucl. Phys. B276 (1986) 241;
D. Espriu, E. de Rafael and J. Taron, Nucl. Phys. B345 (1990) 22;
H. Georgi, preprint HUTP-92/A036.

\bibitem{Hidden}
A.P. Balachandran, A. Stern and G, Trahern, Phys. Rev. D19 (1979) 2416;
M. Bando, T. Kugo  and K. Yamawaki, Phys. Rep. 164 (1988) 217.

\bibitem{Bando}
M. Bando, T. Kugo, S. Uehara, K. Yamawaki and T. Yanagida,
Phys. Rev. Lett. 54 (1985) 1215; see also G. Ecker, J. Gasser, A. Pich
and E. de Rafael, Nucl. Phys. B231 (1989) 311.

\bibitem{Casalbuoni}
R. Casalbuoni, S. De Curtis, D. Dominici and R. Gatto,
Phys. Lett. B155 (1985) 95; Nucl. Phys. B282 (1987) 235.

\bibitem{Herrero}
R. Casalbuoni, P. Chiappetta, D. Dominici, F. Feruglio and R. Gatto,
Nucl. Phys. B310 (1988) 181;
A. Dobado and M. J. Herrero, Phys. Lett. B228 (1989) 495.

\bibitem{Porrati}
S. Ferrara, A. Masiero and M. Porrati, preprint CERN-TH.6726/92
and NYU-TH-92/11/02.

\bibitem{sugra}
E. Cremmer, B. Julia, J. Scherk, S. Ferrara, L. Girardello and
P. van Niewenhuizen, Nucl. Phys.  B147 (1979);
E. Cremmer, S. Ferrara, L. Girardello and A van Proeyen,
Nucl. Phys. B212 (1983) 413.

\bibitem{superstring}
L. Dixon, V. Kaplunovsky and J. Louis, Nucl. Phys. B355 (1991) 649;
J-P. Deredinger, S. Ferrara, C. Kounnas and F. Zwirner,
Nucl. Phys. B372 (1992) 145 and references therein.

\bibitem{Callan}
S. Coleman, J. Wess and B. Zumino, Phys. Rev. 177 (1969) 2239;
C. Callan, S. Coleman, J. Wess and B. Zumino, Phys. Rev. 177 (1969) 2247.

\bibitem{Haag}
R. Haag, Phys. Rev. 112 (1958) 669;
D. Ruelle, Helv. Phys. Acta 35 (1962) 34;
H.J. Borchers, Nuovo Cim. 25 (1960) 270.

\bibitem{Coleman}
S. Coleman in {Hadron and Their Interactions}, ed. A. Zichichi (Academic
Press. Inc.,New York, 1968)

\bibitem{Higgs}
L. Rolandi, talk given at the XXVI Int. Conf. on High Energy Physics,
Dallas, Aug. 1992,
to appear in the proceedings.

\bibitem{Peccei}
R.D. Peccei and X. Zhang, Nucl. Phys. B337 (1990) 269;
R.D. Peccei, S. Peris and X. Zhang, Nucl. Phys. B349 (1991) 305;
M. Frigeni, R. Rattazzi, Phys. Lett. B269 (1991) 412.

\bibitem{Peskin}
M. Peskin and T. Takeuchi, Phys. Rev. Lett. 65 (1990) 964;
W. Marciano and J. Rosner, Phys. Rev. Lett. 65 (1990) 2963;
D. Kennedy and P. Langacker, Phys. Rev. Lett. 65 (1990) 2967;
A. Ali and G. Degrassi, preprint DESY-91-035;
A. Dobado, D. Espriu, and M.J. Herrero, Phys. Lett. B255 (1991) 405;
R.D. Peccei and S. Peris, Phys. Rev. D44 (1991) 809;
J. Ellis, G.L. Fogli and E. Lisi, preprint CERN-TH-6363-92, BARI-TH-96-92.

\bibitem{AB}
G. Altarelli and R. Barbieri, Phys. Lett. 253B (1991) 161;

\bibitem{ABJ}
G. Altarelli, R. Barbieri and S. Jadach, Nucl. Phys. B369 (1992) 3.

\bibitem{Yellow}
See for instance:
M. Consoli, F. Jagerlehner and W. Hollik in
{Z Physics at LEP 1}, Vol. 1, eds. G. Altarelli, R. Kleiss and C. Verzegnassi,
CERN 89-08, Geneva, 1989

\bibitem{Altarelli}
G. Altarelli, talk given at the Rencontres de Moriond on Electroweak
Interactions and Unified Theories, Les Arcs, France, March 15-22, 1992.

\bibitem{Georgi1}
H. Georgi, Nucl. Phys. B361 (1991) 339

\bibitem{Buch}
W. Buchm\"uller and D. Wyler, Nucl. Phys. B268 (1986) 621

\bibitem{Hagi}
K. Hagiwara, R.D. Peccei, D. Zeppenfeld and K. Hikasa,
Nucl. Phys. B282 (1987) 253.

\bibitem{Zeppe}
K. Hagiwara, S. Ishihara, R. Szalapski and D. Zeppenfeld,
Phys. Lett. B283 (1992) 353;

\bibitem{Valencia}
G. Valencia, talk given at the XXVI Int. Conf. on High Energy Physics,
Dallas, Aug. 1992, FERMILAB-CONF-92/246-T

\bibitem{Bertolini}
S. Bertolini and A. Sirlin, Nucl. Phys. B248 (1984) 589

\bibitem{Bagger}
J.A. Bagger,
Lectures given at the 1991 Theoretical Advanced Studies Institute, Boulder,
Colorado, preprint JHU-TIPAC-910038.

\end{thebibliography}
\end{document}